\documentclass[a4paper,fleqn,usenatbib]{mnras}

\usepackage{newtxtext,newtxmath}

\usepackage[T1]{fontenc}
\usepackage{ae,aecompl}

\usepackage{graphicx}	
\usepackage{amsmath}	
\usepackage{amssymb}	

\title[3D asymmetrical kinematics of mono-age populations]{Mapping the Galactic disk with the LAMOST and Gaia Red clump sample. II. 3D asymmetrical kinematics of mono-age populations in the disk between 6$-$14 \,kpc}

\author[H.-F. Wang et al.]{H.-F. Wang$^{1,2,3}$\thanks{E-mail: hfwang@bao.ac.cn(HFW)}, 
M. L\'opez-Corredoira$^{5,6}$,
Y. Huang$^{1}$, 
J. L. Carlin$^{4}$, 
B.-Q. Chen$^{1}$, 
\newauthor{C. Wang$^{7,8,3}$, J. Chang$^{9,10}$,
H.-W. Zhang$^{7,8}$, 
M.-S. Xiang$^{11,9}$,
H.-B. Yuan$^{12}$, 
W.-X. Sun$^{1}$, 
\newauthor{X.-Y. Li$^{1}$, 
Y. Yang$^{1}$, 
L.-C. Deng$^{9,2}$}}
\newauthor
\\
$^{1}$South$-$Western Institute for Astronomy  Research, Yunnan University, Kunming, 650500, P.\,R.\,China \\
$^{2}$Department of Astronomy, China West Normal University, Nanchong 637009, China \\
$^{3}$LAMOST Fellow \\
$^{4}$LSST, 950 North Cherry Avenue, Tucson, AZ 85719, USA \\
$^{5}$Instituto de Astrof\'\i sica de Canarias, E-38205 La Laguna, Tenerife, Spain \\
$^{6}$Departamento de Astrof\'\i sica, Universidad de La Laguna, E-38206 La Laguna, Tenerife, Spain \\
$^{7}$Department of Astronomy, Peking University, Beijing 100871, People's Republic of China \\
$^{8}$Kavli Institute for Astronomy and Astrophysics, Peking University, Beijing 100871, People's Republic of China \\
$^{9}$Key Laboratory of Optical Astronomy, National Astronomical Observatories, Chinese Academy of Sciences, Beijing 100012, \\
People's Republic of China \\
$^{10}$Purple Mountain Observatory, the Partner Group of MPI f$\ddot{u}$r Astronomie, 2 West Beijing Road, Nanjing 210008, China \\
$^{11}$Max-Planck Institute for Astronomy, Konigstuhl, D-69117, Heidelberg, Germany\\
$^{12}$Department of Astronomy, Beijing Normal University, Beijing 100875, People's Republic of China}

\date{Accepted XXX. Received YYY; in original form ZZZ}

\pubyear{2019}

\begin{document}
\label{firstpage}
\pagerange{\pageref{firstpage}--\pageref{lastpage}}


\begin{abstract}	

We perform analysis of the three-dimensional kinematics of Milky Way disk stars in mono-age populations. We focus on stars between Galactocentric distances of $R=6$ and 14 \,kpc, selected from the combined LAMOST DR4 red clump giant stars and Gaia DR2 proper motion catalogue. We confirm the 3D asymmetrical motions of recent works and provide time tagging of the  Galactic outer disk asymmetrical motions near the anticenter direction out to Galactocentric distances of 14\,kpc. Radial Galactocentric motions reach values up to 10 km s$^{-1}$, depending on the age of the population, and present a north-south asymmetry in the region corresponding to density and velocity substructures that were sensitive to the perturbations in the early 6 \,Gyr. After that time, the disk stars in this asymmetrical  structure have become kinematically hotter, and are thus not sensitive to perturbations, and we find the structure is a relatively younger population. With quantitative analysis, we find stars both above and below the plane at $R\gtrsim 9$ kpc that exhibit bending mode motions of which the sensitive duration is around 8 \,Gyr.  We speculate that the in-plane asymmetries might not be mainly caused by a fast rotating bar, intrinsically elliptical outer disk, secular expansion of the disk, or streams. Spiral arm dynamics, out-of-equilibrium models, minor mergers or others are important contributors. Vertical motions might be dominated by bending and breathing modes induced by complicated inner or external perturbers. It is likely that many of these mechanisms are coupled together.

\end{abstract}
 
\section{Introduction}
With modern large scale Galactic surveys, Galactoseismology is starting to become reality, in which we can unravel the chemo-dynamical history of the disk, and in turn infer its history of accretion and the characteristics of the satellite galaxies \citep[e.g.,][]{Antoja2018}. \citet{Antoja2018} discovered the Milky Way is full of substructures such as arches, boxy shapes, spirals, snail shells, and ridges, and suggested that these substructures are from phase mixing. Many follow-up studies of disk structure(s) have appeared recently \citep{Tian2018,Wangchun2019, Li2019}.  For example, \citet{Tian2018} used stars in common between LAMOST and Gaia to find that phase-space spirals exist in all the spatial bins, and that the vertical perturbation probably started no later than 0.5 Gyr ago and will disappear in $\sim$6 \,Gyr. This demonstrates that stellar ages can help us to disentangle similar asymmetric structures in the Galactic disk. 

The Milky Way's non-axisymmetric kinematics have also been mapped with data from Gaia DR2 \citep{Katz2018,Lopez2018, wang2019a, wang2019c}. Gaia DR2 provides unprecedentedly accurate measurements of proper motions, positions, parallaxes, and line-of-sight velocities for 7.2 million stars brighter than $G_{RVS}$= 12 \,mag \citep{Lindegren2018}. With this, \citet{Katz2018} obtained detailed 3D kinematics and asymmetries for Galactocentric distances
$R<13$ kpc. Some similar results are also presented by \citet{Wang2018a}, who found that Galactic outer disc stars in the range of Galactocentric distance between R = 8 and 13\, kpc and vertical position between Z = $-$2 and 2 \,kpc exhibit asymmetrical motions in the Galactocentric radial, azimuthal, and vertical directions. 
\citet{Carrillo2019}, assuming priors about the stellar distribution, extended Gaia-DR2 kinematic maps up to $R=14-16$ \,kpc. In \citet{Lopez2018}, using a statistical deconvolution  of the parallax errors, the kinematics maps of Gaia-DR2 data were extended up to $R=20$ \,kpc.

In-plane non-axisymmetries in the Milky Way disk kinematics have been shown in many works \citep[e.g.,][]{Siebert11, Antoja12, Siebert12, Lop141, Xia15, Liu171, Lop16, Tian172}. For example, \citet{Siebert11} found a radial velocity gradient in the direction of the Galactic center by using RAVE \citep{Kunder17} red clump stars. Many diagonal ridge features were found in the R $-$ $V_{rot}$ map, the locations of which were compared with the locations of the spiral arms and the expected outer Lindblad resonance of the Galactic bar \citep{Kawata2018}.  Some possible mechanisms such as  the Galactic bar's outer Lindblad resonance, perturbations due to the bar or spiral arms \citep {Denhen00,Fux01,Quillen05}, by external minor mergers such as the Sagittarius dwarf galaxy passing by, or by interaction with the Magellanic Clouds \citep{Minchev09, Minchev10, Gomez121, Gomez122} are proposed for these in-plane non-axisymmetries.

Vertical non-axisymmetries  and wave-like density patterns are found in the solar neighborhood \citep{Widrow12, Williams13} and in the outer disk \citep{Xu15, Wang2018b}. One example is the clear asymmetrical overdensity structure around $Z\approx$ 0.5 \,kpc,  $R\approx$ 10$-$11 \,kpc, $b \approx$ 15 degree, which is called the north near structure in \citet{Xu15}. This structure is located around 2 \,kpc from the Sun. \citet{Xiang2018} also revealed the density asymmetries in mono-age populations with main sequence turn-off stars. Besides these density asymmetries, many velocity asymmetries have been identified in the Milky Way disk \citep{Carlin13, Sun2015, Carrillo17, Pearl17, Wang2018a}. Scenarios for producing these structures include minor mergers or interactions with nearby dwarfs or satellites \citep{Gomez13, Donghia16, Laporte18}. The effects of even lower-mass dark matter subhalos have also been invoked as  a possible explanation \citep{Widrow14}. The Galactic outer disk warp can also excite vertical bulk motions \citep{roskar10,Lop141}. The warp may be generated by the interaction with the Magellanic Clouds \citep{Burke1957, Weinberg2006},  interaction with Sagittarius \citep{Bailin2003a}, disk bending instabilities \citep{Revaz2004}, misaligned gas infall \citep{Ostriker1989, Quinn1992, Bailin2003b}, intergalactic magnetic fields \citep{Bat98}, or accretion of the intergalactic medium directly onto the disc \citep{Lop02a}. The kinematical signature of vertical bulk motions can also be used to constrain the warp's properties.

Recently, we have found many outer disk density asymmetries \citep{Wang2018b} and investigated the velocity asymmetrical motions as far away as 5 \,kpc from the solar location at low Galactic latitudes in \citet{Wang2018a}.  Apart from this, much progress has been made in Galacto-seismology with the help of ground-based proper motions \citep[e.g.,][]{Siebert11,Widrow12, Williams13}. The proper motion accuracies provided by Gaia have brought us into the golden era of disc kinematics studies. However, stellar ages and abundances from large spectroscopic surveys are needed for us to assess which mechanisms may have created these asymmetries and how they evolved.
 The LAMOST spectroscopic survey \citep{Cui12, Zhao12, Deng2012, liu2014} has now provided a vast sample of red clump giant stars (RCG) with line-of-sight velocities, metallicities, abundances, and ages.  For this paper, we use red clump giants with estimated ages and distances, combined with the Gaia DR2 proper motion catalogue, to decipher the outer disk 3-dimensional kinematical structure and asymmetries in more detail, focusing especially on patterns in radial and vertical velocities with age. 

The paper is organized as follows. Section 2 describes the sample and coordinate transformations. Section 3 shows the reconstruction of recent Gaia work. The radial and vertical velocity distribution projected onto 2D maps in the different mono-age populations, and the distributions of 3-dimensional velocities along the radial directions are shown in Section 4. We also introduce the vertical asymmetrical motions including some quantitive analysis. In Section 5,  we discuss the asymmetrical mechanisms and some validations qualitatively.  A summary is given in Section 6.

\section{Data}
\subsection{Sample selection and Distance}
The Large Aperture Multi-Object Fiber Spectroscopic Telescope (LAMOST, also called the Guo Shou Jing Telescope), is a quasi$-$meridian reflecting Schmidt telescope with an effective aperture of about 4 meters. A total of 4000 fibers, capable of obtaining low resolution spectra (R $\sim$ 1800) covering the range from 380 to 900 nm simultaneously, are installed on its $5^{\circ}$ focal plane \citep{Cui12, Zhao12, Deng2012, liu2014}. The LAMOST Data Release 4 (DR4) red clump catalogue contains more than 0.15 million spectra, with most of them being primary red clump stars, and a few percent secondary RCGs and Red Giant Branch (RGB) stars. LAMOST catalogs contain reliable stellar parameters, radial velocity, and [$\alpha$/Fe], and derived quantities including distances and ages using the Kernel Principal Component Analysis (KPCA) method \citep{Huang2019}. The sample is mainly distributed in the Galactic Anti-Center direction due to special conditions at the site \citep{Yao12}.  We select red clump stars for this analysis because they are reliable standard candles in the disk \citep{Huang2019} and the method for the determination of stellar parameters, metallicity, abundance and age has been tested extensively \citep{Xiang2017a, Xiang2017b, Xiang2017c, wu2018, wu2019}.

\begin{figure}
  \centering
  \includegraphics[width=0.48\textwidth]{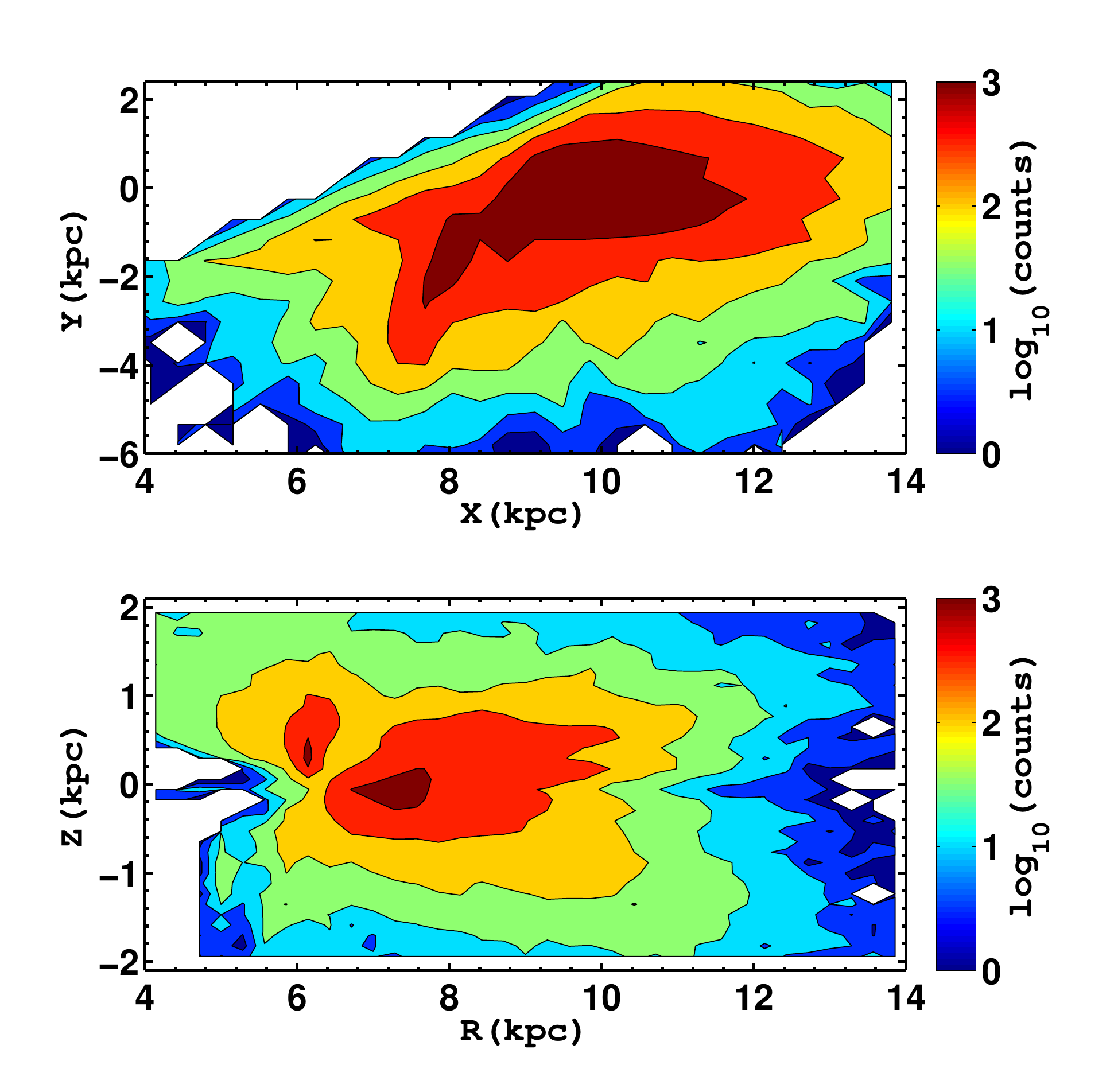}
  \caption{Spatial distribution of the RCG stars used in this work. The top panel is the $X$ and $Y$ in Galactic coordinates, and the bottom panel shows $R$ and $Z$ in cylindrical coordinates separately, color coded by star counts on a log scale. The Sun is at $(X, Y) = (8.34, 0)$~kpc and $(R, Z) = (8.34, 0.027)$~kpc in these panels.}
  \label{rcxyzr}
\end{figure}

Interstellar extinction was derived using the star pairs method \citep{Yuan15}; with the help of large sky area spectroscopic surveys, observations of stars of essentially identical stellar atmospheric parameters in different environments can be easily paired and compared in different regions, and intrinsic colors/color excess of individual stars estimated with accuracies of 1- 4\%. The technique is able to determine E(B-V) to an accuracy of 0.01 mag. Red clump stars are standard candles, with distance uncertainties around 5-10\% \citep{Huang2019}. In this work, we only have primary red clump stars of which details can also be found in \citet{Huangyang2015}. Fig.~\ref{rcxyzr} shows spatial distribution of the RCG stars used in this work. The top panel is looking down on the Galactic plane at Galactic $X$ and $Y$ coordinates, and shows that our sample is mainly distributed in the anti-center direction. The bottom panel of Fig.~\ref{rcxyzr} is the distribution in $R$ and $Z$ Galactic cylindrical coordinates, color coded by star counts on a log scale. The majority of stars in our sample are outside the Solar radius in the direction of the low latitude Galactic Anticenter, with higher sampling rates in the North Galactic Cap. The coordinates used throughout this work have $X$ increasing outward from the Galactic centre, $Y$ in the direction of rotation, and $Z$ positive towards the North Galactic Pole (NGP) \citep{Williams13}.

\subsection{3D Velocities derivation }

The Gaia DR2 catalogue contains high-precision positions, parallaxes, and proper motions for 1.3 billion sources as well as line-of-sight velocities for 7.2 million stars. For stars of G$<$14 \,mag, the median uncertainty  is 0.03 \,mas for the parallax and 0.07~mas~yr$^{-1}$ for the proper motions. For the stars with available heliocentric radial velocities (7\,224\,631 sources), the  parallax error is less than 100$\%$ for 7\,103\,123 of them. The Radial Velocity Spectrometer collects medium-resolution spectra with $R\gtrsim11700$ which cover the wavelength range 845-872 nm centered on the calcium triplet region \citep{Gaia2018a, Katz2018, Lopez2018}.  The sources with radial velocities are nominally brighter than 12 \,mag in the $G_{RVS}$ photometric band. As described in \citet{Wang2018a}, we derive the 3D velocities assuming the location of Sun is $R_{\odot}$  = 8.34 \,kpc \citep{Reid14} and $Z_{\odot}$ = 27 \,pc \citep{Chen01}. The heliocentric rectangular components of the Galactic space velocity $U, V$, and $W$ are determined by the right-handed coordinate system based on \citet{Johnson87}, with $U$ positive towards the Galactic centre, $V$ positive in the direction of Galactic rotation, and $W$ positive towards the north Galactic pole. 

For the solar motion we use the \citet{Tian15} values: [$U_{\odot}$, $V_{\odot}$, $W_{\odot}$] = [9.58, 10.52, 7.01] km s$^{-1} $. We also ran our analysis with other solar motions \citep[e.g., ][]{Huang2015}, and found that the results are still robust. The circular speed of the LSR is adopted as 238 km s$^{-1} $ \citep{Schonrich12}.  We also correct the radial velocities of the samples by adding 3.1 km s$^{-1}$ in this work to place them on the APOGEE velocity scale \citep{Blanton17}.

\begin{figure}
  \centering
  \includegraphics[width=0.5\textwidth]{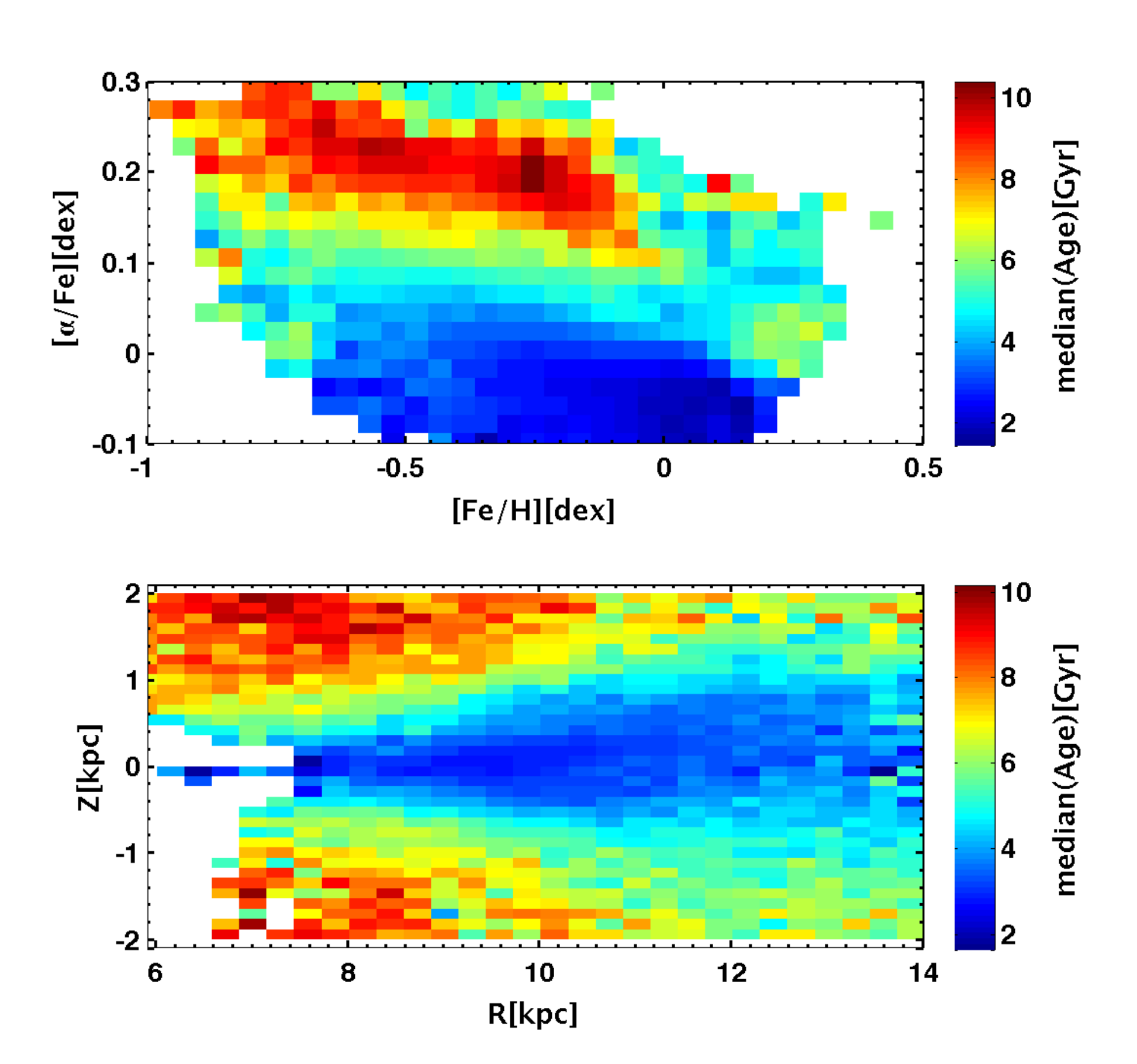}
  \caption{The top panel shows the age distribution of red clump giant stars on the [Fe/H] and [$\alpha$/Fe] plane. The color coding corresponds to the age for each bin. The bottom panel shows the median age of stars in the $R, Z$ plane in the range of $Z=[-2, 2]$ \,kpc, $R=[6, 14]$ \,kpc. There is a clear thin disk, thick disk, and flaring disk population according to chemistry and age definitions of \citet{Martig2016}.}
  \label{fehageaferz}
\end{figure}

The behavior of alpha abundance ([$\alpha$/Fe]), metallicity ([Fe/H]), and age measurements for red clump giant stars within the range of $Z = [-2, 2]$ \,kpc is shown in Figure~\ref{fehageaferz}, showing a color-coded map of the median age (top panel) and its associated error distribution (bottom) in the $R$ (radial distance from the Galactic centre), $Z$ (vertical distance from the midplane) plane. The median number of stars per pixel in both figures is larger than 50. The top panel shows that the age has a clear gradient from thin disk to thick disk population from the chemical distribution, in agreement with the classifications of \citet{Martig2016}. There is a young [$\alpha$/Fe] enriched population in the top right corner. This intriguing population will be discussed in \citet{Sun2019}. Errors in age and [$\alpha$/Fe] could contribute to this region, as suggested by the one red strip of old $\alpha$-enhanced stars that appears to be superimposed on an extended (light blue) background. The typical error of [$\alpha$/Fe] is around 0.05, the age error is around 30\%. Even after accounting for these errors, many stars are still strange for the current chemical evolution models; we assume that most of the stars with relatively young age but high [$\alpha$/Fe] are real for the current work. More tests and details about this extended background will be detailed in \citet{Sun2019}. 
 The bottom panel of Figure~\ref{fehageaferz} shows the age distribution in the $R, Z$ plane. There are clear disk flaring features around $|Z|=0.8$ \,kpc. Some discussions of disk flaring mechanisms can be found in \citet{Quillen93, Solway2012, Bovy2016, Minchev2015, Xiang2018, Minchev2016, Minchev2018}. The bottom panel of this figure shows a clear negative age gradient in the morphological thick disk, as predicted by \citet{Minchev2015} and later shown in APOGEE \citet{Martig2016}.


For our sample during this paper, the stars located inside $|Z|$ $<$ 3 kpc and 4 $<$ R $<$ 14 \,kpc are selected to map the lower disk kinematics, and we mainly focus on the region within $Z=[-2, 2], R=[6, 14]$ \,kpc,  where we have small random and systematic errors. The stars with LAMOST spectroscopic SNR $<$ 20 and derived age larger than 15 \,Gyr are not included.  A few hundred stars, distributed randomly, exhibit relative errors on their ages larger than 50$\%$. In this work we do not use relative errors to exclude stars to reduce Poisson error. In order to reduce the contamination of halo stars, we only use stars with [Fe/H] greater than -1.0 \,dex, so we can focus on Galactic disk populations. We also set some criteria in velocity to remove fast-moving halo stars: $V_R$=[-150, 150] km s$^{-1}$, $V_{\theta}$=[-50, 350] km s$^{-1}$, and $V_Z$=[-150, 150] km s$^{-1}$; this removes a small fraction of stars.

\section{Reconstructing Gaia DR2 work}   

\begin{figure}
  \centering
  \includegraphics[width=0.5\textwidth]{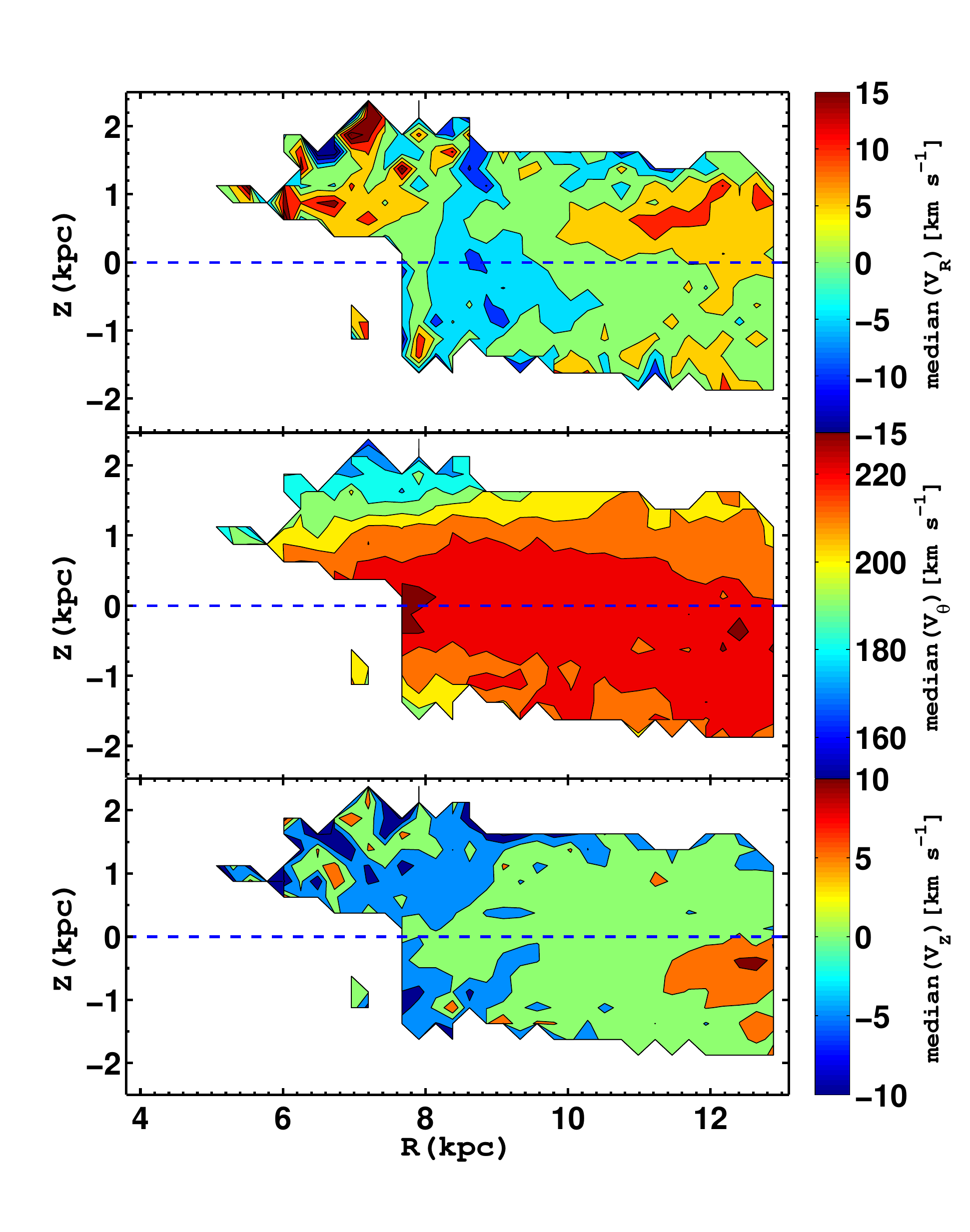}
  \caption{Edge$-$on views of the kinematics of the disk with similar range of \citet{Katz2018}, derived using the giant sample: from top to bottom, median velocity maps of $V_{R}$, $V_{\theta}$, $V_{Z}$(in km$^{-1}$). Each bin in the maps contains at least 50 stars.}
  \label{recon_katz}
\end{figure}

\begin{figure}
  \centering
  \vspace{-2.5cm}
  \includegraphics[scale=0.48]{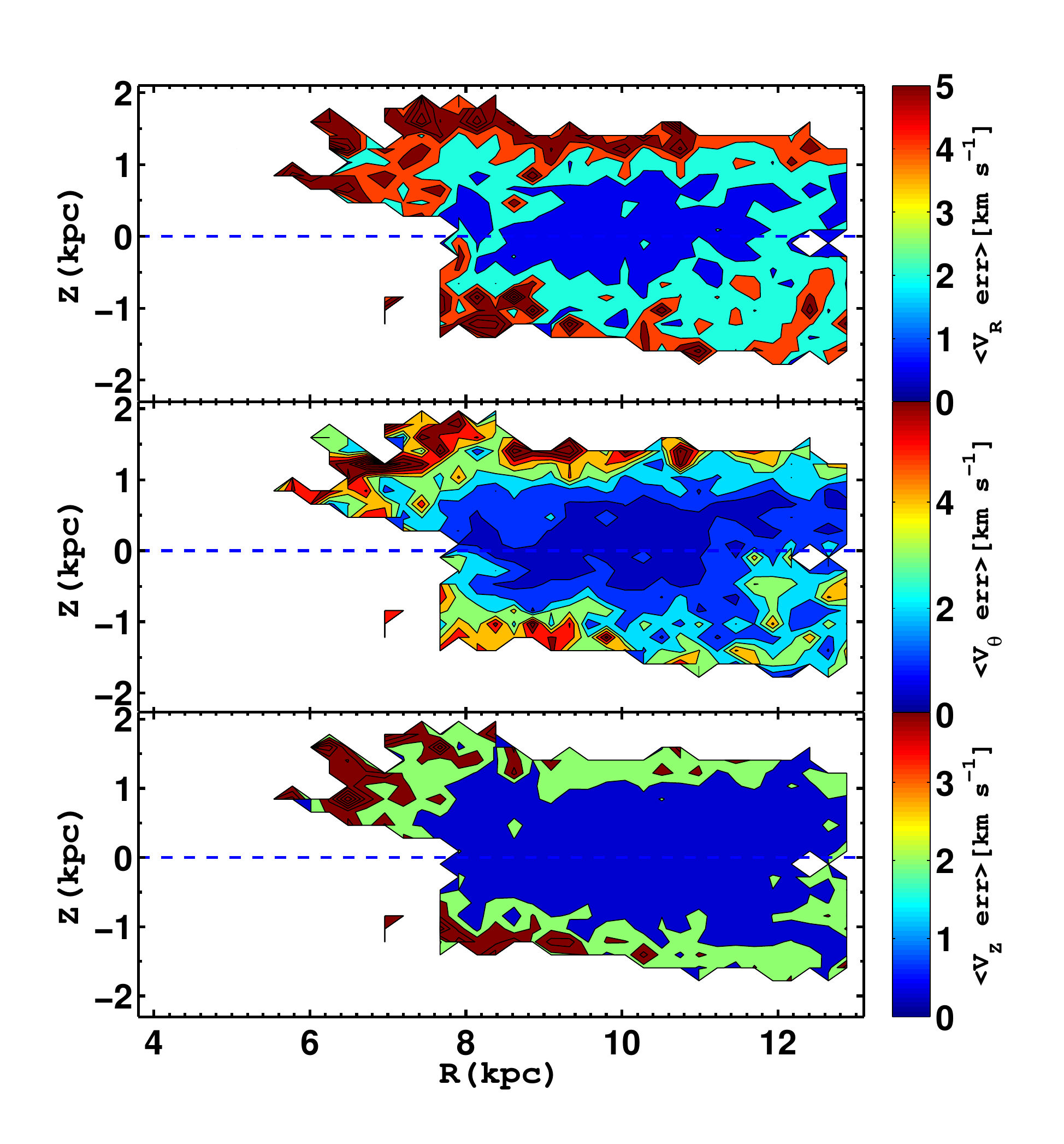}
  \caption{Error analysis for Fig.~\ref{recon_katz}, with each panel showing bootstrap errors for the corresponding quantities in Fig.~\ref{recon_katz}. In this figure, the line-of-sight velocity is from LAMOST and the proper motion is from Gaia.}
  \label{recon_katz_ERROR}
\end{figure}

We reconstruct recent work by \citet{Katz2018} using stars in common between LAMOST and Gaia in Fig.~\ref{recon_katz}. The top panel shows some oscillations in the northern (0 $<$ Z $<$ 2 kpc) stars, a well-known nearby northern structure in the velocity field around R$\sim$10$-$12 and Z$\sim$0.5 \,kpc at a Galactic latitude of around 15 degrees. The middle panel shows an outward flaring or displacement of the contours away from $Z=0$. This structure can be contributed by an increase in asymmetric drift, or we can say it is attributed by stars on more energetic orbits reaching the higher-Z region, which will necessarily have lower angular momentum.  The latter description comes to the same thing dynamically but would be more direct. In \citet{Wang2018a} we call it asymmetrical structure with the contribution of asymmetric drift. The bottom panel shows stars in the north side and south side of the plane at R$>$9 kpc exhibit net upward vertical motions larger than 2 km~s$^{-1}$ if we check the color in the range of -5 to 5 km s$^{-1}$. This is similar to results shown by \citet{Wang2018a}; it is a bending mode \citep{Widrow12, Widrow14,Chequers18} or vertical upward bulk motions. Comparing to \citet{Katz2018} in more detail, we note that in Figure~\ref{vz_age_benbrea}, the bending pattern seen as redder colored bins is clear and stronger. Fig.~\ref{recon_katz_ERROR} shows the error analysis for Fig.~\ref{recon_katz} results, corresponding to bootstrap errors on $V_{R}$,$V_{\theta}$, $V_{Z}$ (in km s$^{-1}$). Bootstrap errors are determined by resampling (with replacement) 100 times for each bin, and the uncertainties of the estimates are determined using 15th and 85th percentiles of the bootstrap samples. In this figure, the line-of-sight velocity is from LAMOST and the proper motion is from Gaia. We can see there are small errors around a few km s$^{-1}$ in each panel.

As we can see, we can reconstruct the Gaia recent work and we can see clear asymmetrical features in the 3D velocity distributions. Apart from contributions to the rotational velocity by asymmetric drift, we also see the clear north near asymmetric structure and vertical asymmetrical motions. In the next section, we will give the time tagging on the radial and vertical asymmetries by using red clump ages.

\section{Time tagging the asymmetrical structure in mono$-$age populations}   

\subsection{Time tagging the north near structure using mono$-$age populations}   

\begin{figure*}
  \centering
  \includegraphics[width=1.0\textwidth]{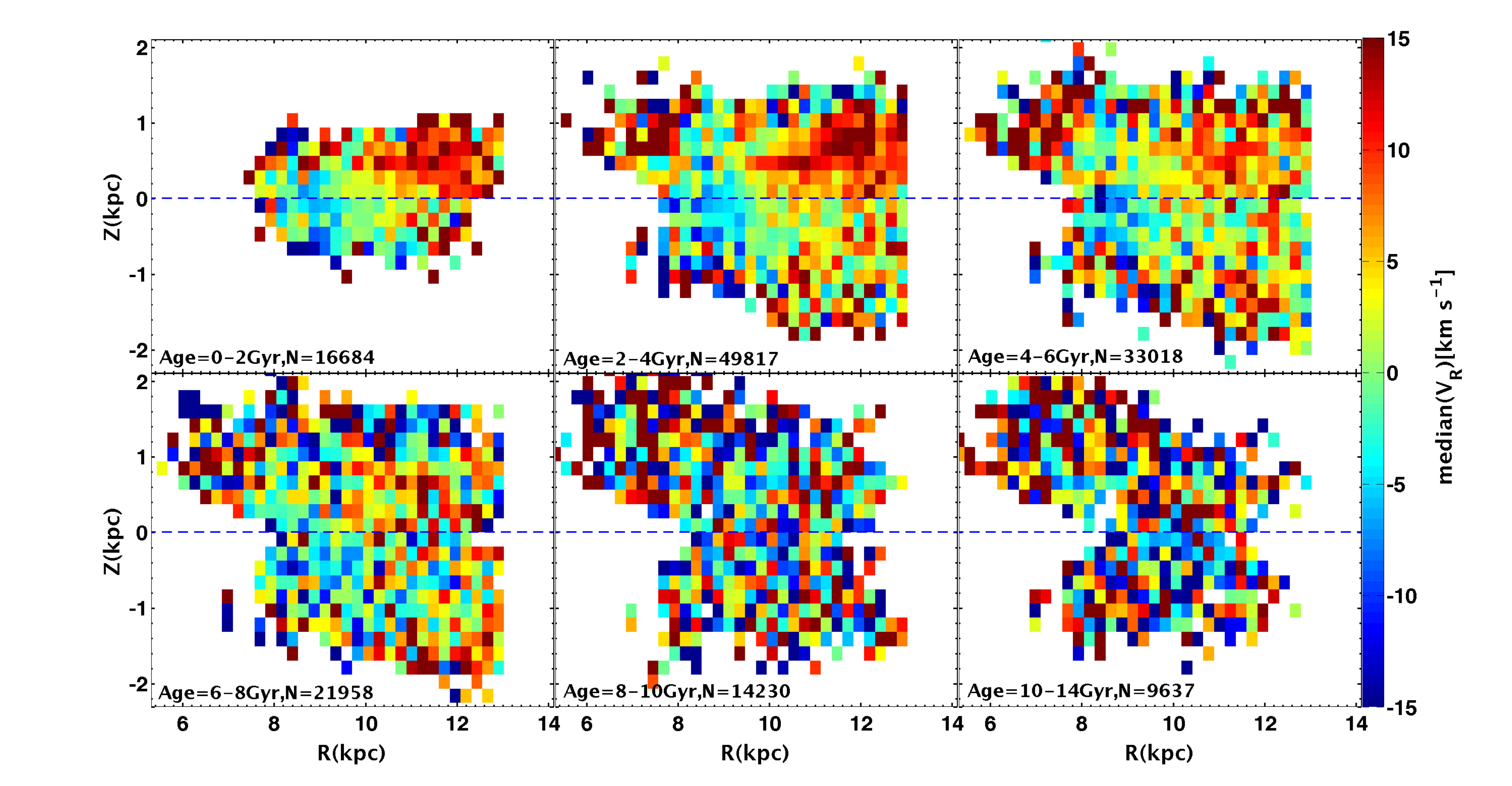}
  \caption{The radial asymmetrical structures in the R, Z plane of the LAMOST-Gaia stars in different age populations. Each panel is colored as median velocity in different age bins. The top panel has clear asymmetries from 0-6 \,Gyr. Each pixel plotted in this figure has at least 10 stars.}
  \label{vR_age_benbrea}
\end{figure*}

\begin{figure*}
  \centering
  \includegraphics[width=1.0\textwidth]{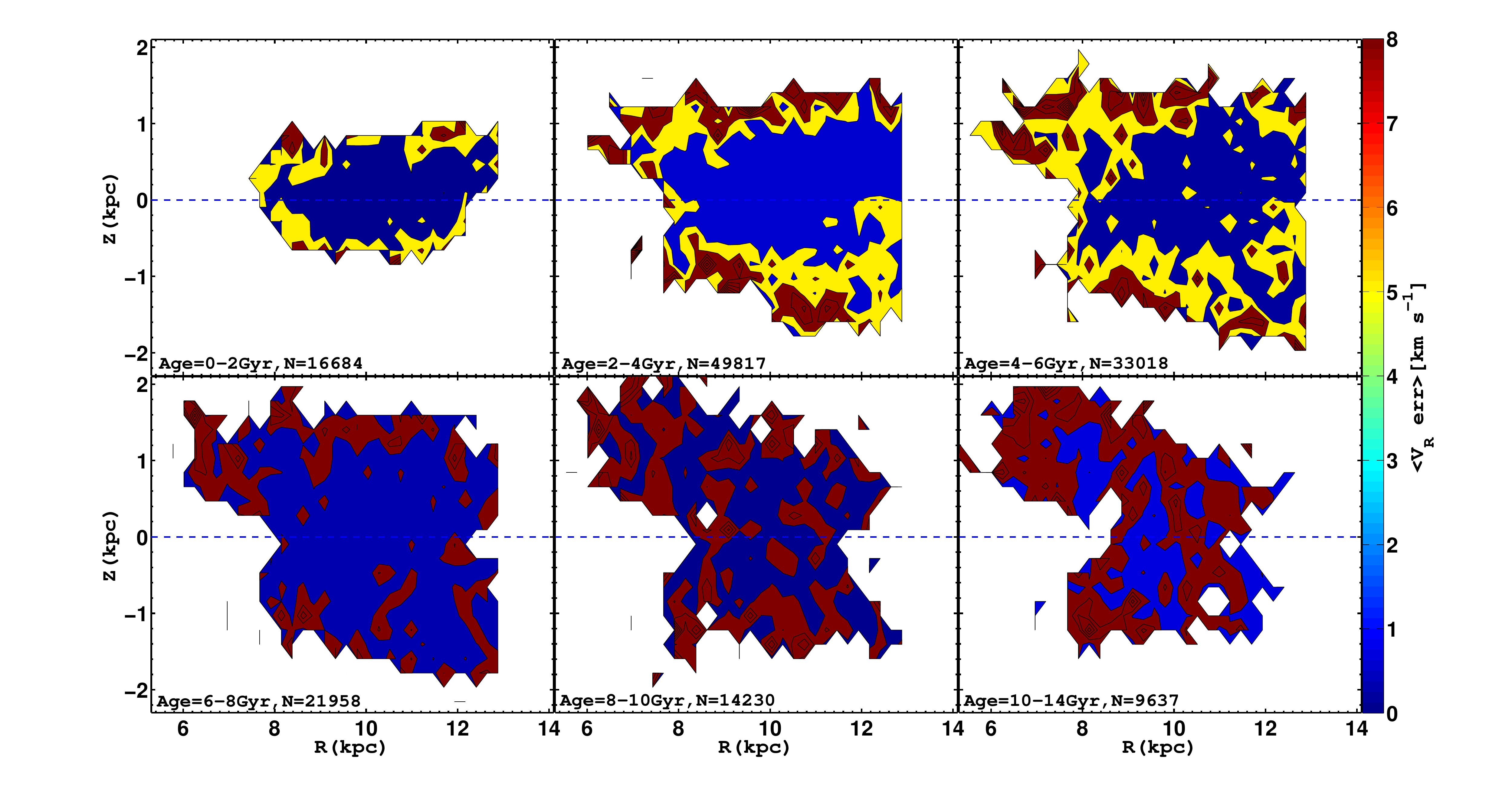}
  \caption{Error analysis of the radial asymmetrical structures in the R, Z plane  for Fig.~\ref{vR_age_benbrea}, with each panel showing bootstrap errors for the corresponding quantities in Fig.~\ref{vR_age_benbrea}.}
  \label{vR_age_benbrea_error}
\end{figure*}

\begin{figure}
  \centering
  \includegraphics[width=0.48\textwidth]{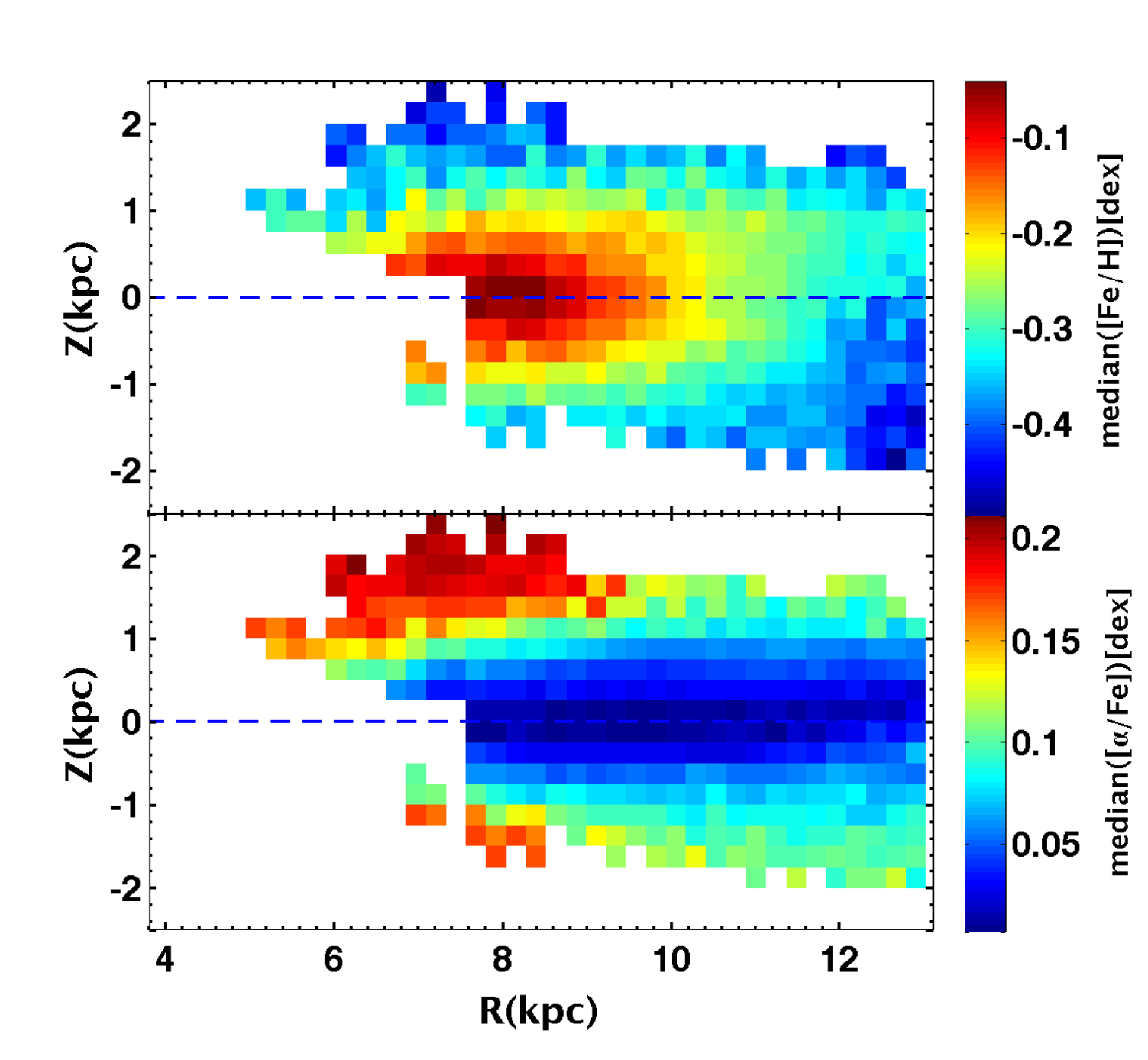}
  \caption{This figure shows the metallicity distribution function and abundance distribution function; there are clear gradients for thin disk and thick disk stars in the R, Z plane. This figure's axes correspond to Fig.~\ref{recon_katz}.}
  \label{rcrzfehafe}
\end{figure}

In this section, we give time stamps for describing 2D asymmetrical radial motions in the Galactic disk.
Fig.~\ref{vR_age_benbrea} shows the variation of Galactocentric radial velocity $V_R$ with age in the $R, Z$ plane, with the numbers of stars used to produce the maps are given in each panel. We can see the large prominent velocity structure on the northern ($0 < Z < 2$ \,kpc) side from $9 < R \lesssim 12$~kpc in the first three panels with velocity clearly larger than 10 \, km s$^{-1}$, in which $V_R$ has a positive gradient and becomes stronger. We also note that there is a north-south asymmetry at the location of $R \sim$ 9-12 \,kpc, $Z \sim$ 0.5 \,kpc with redder bins named and confirmed as the north near substructure in \citet{Wang2018a}. From this figure, we can see that this feature disappears around 6-8 \,Gyr, which implies that the time when the north near substructure was sensitive to the perturbers is around 6 \,Gyr. According to the error analysis with bootstrap method shown in Fig.~\ref{vR_age_benbrea_error}, we can see that most of the error values are around 1-2 \, km s$^{-1}$ for the first four panels, when the age is larger than 8 \,Gyr. The error of the last two panels is significantly larger, but it cannot change our final conclusion of being sensitive to time, which is mainly based on the first four figures of Fig.~\ref{vR_age_benbrea_error}. We also think it is a relatively younger or medium age structure according to the age distribution of corresponding regions in Fig.~\ref{fehageaferz}. The north near structure region's age is less than 6 \,Gyr and most of stars are around 2 \,Gyr.  Meanwhile, [$\alpha/Fe$] and [Fe/H] is also showing that this structure is a low $\alpha$ and metal rich population of stars as shown in Fig.~\ref{rcrzfehafe}, in which there also is a clear gradient for thin disk and thick disk. Fig.~\ref{rcrzfehafe} (lower panel) shows the high-$\alpha$ population extending only to about R = 9\.kpc at Z = 1-2 \,kpc. This important effect was reported by \citet{Hayden2015}. We present a full map of the north near structure with age information thanks the larger sky LAMOST spectroscopic survey. 

\begin{figure*}
  \centering
  \includegraphics[width=0.98\textwidth]{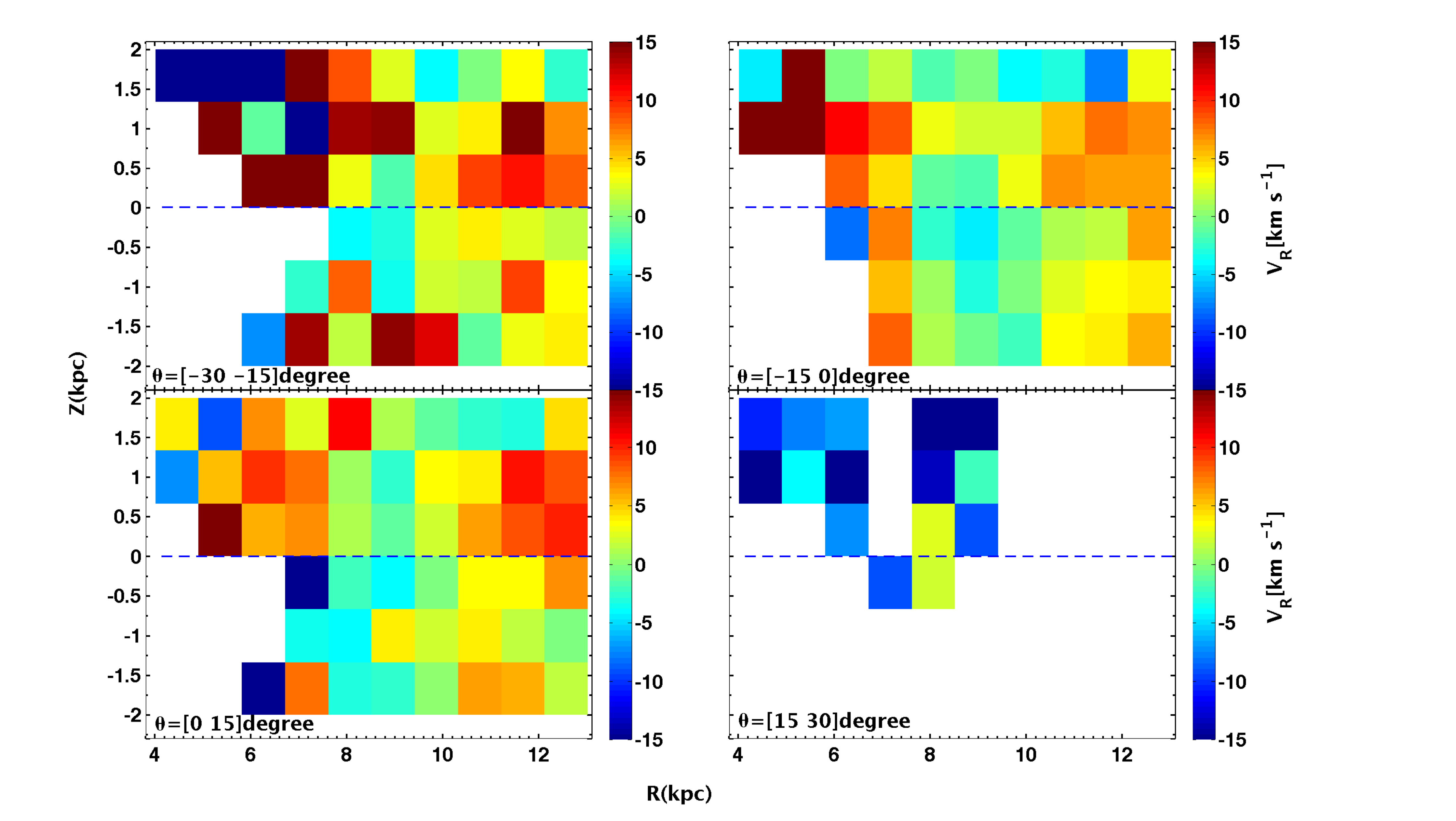}
  \caption{Edge-on maps of the median radial velocity $V_R$ of the red clump sample. Each map corresponds to a slice of 15 degrees in azimuth: [-30, -15] (top left), [-15, 0] (top right), [0, +15] (bottom left), and [+15, +30] degrees (bottom right). The Sun is at X = 8.34 kpc and Y = 0 kpc. The Galactic centre is located on the left side (at $R = 0$~kpc).}
  \label{vr_projectionsi}
\end{figure*}

In order to reduce the influence of projection effects on our results, we show in Fig.~\ref{vr_projectionsi} edge-on maps of the median radial velocity $V_R$ of the red clump sample. Each map corresponds to a slice of 15 degrees in azimuth: [-30, -15] (top left), [-15, 0] (top right), [0, +15] (bottom left), and [+15, +30] degrees (bottom right). The Sun is around X = 8.34 kpc and Y = 0 kpc.  We can see the north near asymmetry is stable in the first three panels at Z$\sim$ 0.5 \,kpc with red color. At $\theta > 15^\circ$, the sampling rates for LAMOST is low so that we cannot see clear patterns. 

Similarly, we also tested the robustness of rotational asymmetrical velocity distribution, which is still stable. However, as we mentioned in section 3, asymmetric drift is contributing to the rotational feature so that we do not make more detailed analysis in this work.

\subsection{Time tagging the vertical motions of mono$-$age populations}   

\begin{figure*}
  \centering
    \includegraphics[width=1.0\textwidth]{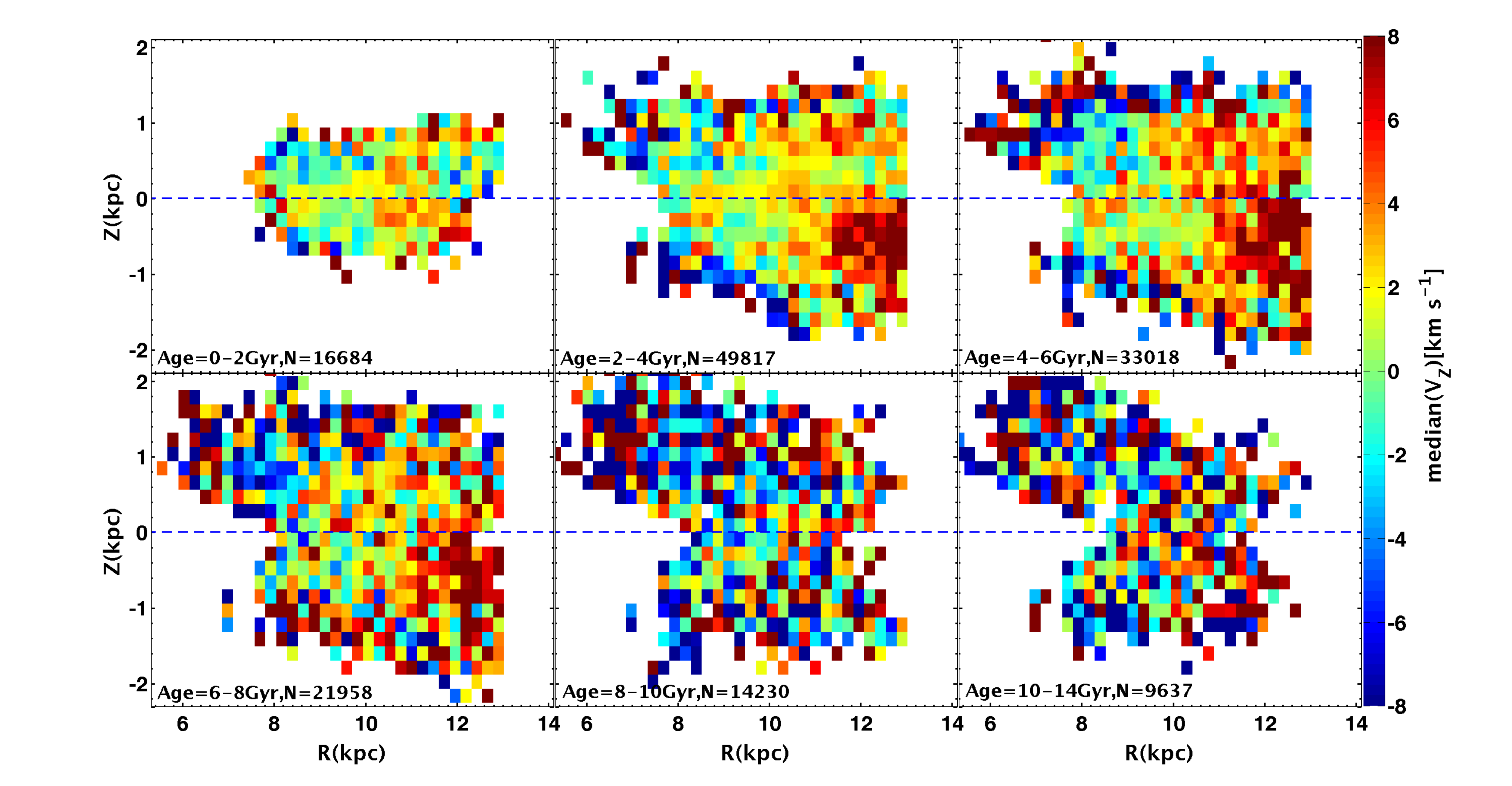}
  \caption{Similar to Fig.~\ref{vR_age_benbrea}, but for vertical velocity, $V_Z$. In the [2-4]\,Gyr population, there are significant bulk motions at almost all radii and heights outside the Sun's radius, extending until $R \sim$ 13 \,kpc. For 8-10 \,Gyr, there are no clear  bulk motion or bending mode features.}
  \label{vz_age_benbrea}
\end{figure*}

\begin{figure*}
  \centering
    \includegraphics[width=1.0\textwidth]{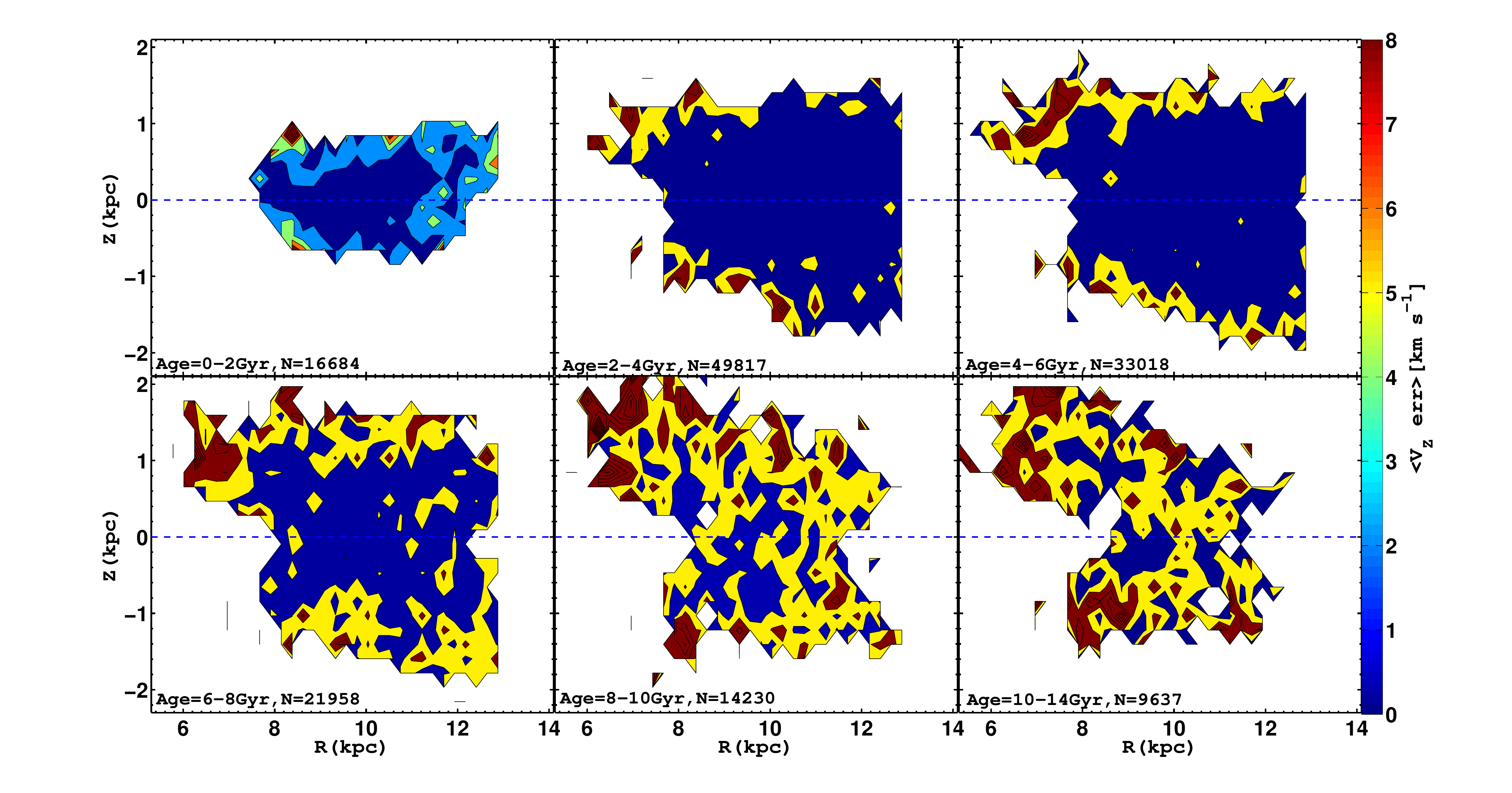}
  \caption{Error analysis of the vertical asymmetrical structures in the R, Z plane  for Fig.~\ref{vz_age_benbrea}, with each panel showing bootstrap errors for the corresponding quantities in Fig.~\ref{vz_age_benbrea}.}
  \label{vz_age_benbrea_error}
\end{figure*}

Figure~\ref{vz_age_benbrea} shows the average $V_{Z}$ with position in the $R, Z$ plane in different age populations. There is clear evidence of a vertical bulk motion outside the solar radius or 9 \,kpc for age in the range of 2-6 \,Gyr: above the midplane, the overall trend shows stars are moving upward, and below the plane, stars are also moving upward. This is very similar to the substructure found by \citet{Wang2018a}. This illustrates what is often called a ``bending'' mode. For 0-2 or 6-8 \,Gyr, there is also a bulk motion or bending mode features with many bins larger than 6  km s$^{-1}$ , but not as strong and clear as in the 4-6\,Gyr population. This bending mode disappears when the age is older than 8 \,Gyr, so we speculate these bulk motions might be sensitive only in the earliest 8 \,Gyr.  After that time, the disk stars for this region become older and kinematically hotter and not sensitive to the possible perturbations. The error analysis in the R, Z plane is given in Figure~\ref{vz_age_benbrea_error}. As is shown with colors, most of the errors in the first four figures are around 1 \,km s$^{-1}$, with larger errors in the last two figures due to the contributions of age error, Poisson error, and velocity error. However, most of the error value of the fifth figure is around 1-5 \,km s$^{-1}$. We do not think it will change our current conclusion of time sensitivity by comparing the values of Figure~\ref{vz_age_benbrea} carefully.

\begin{figure*}
  \centering
  \includegraphics[width=0.98\textwidth]{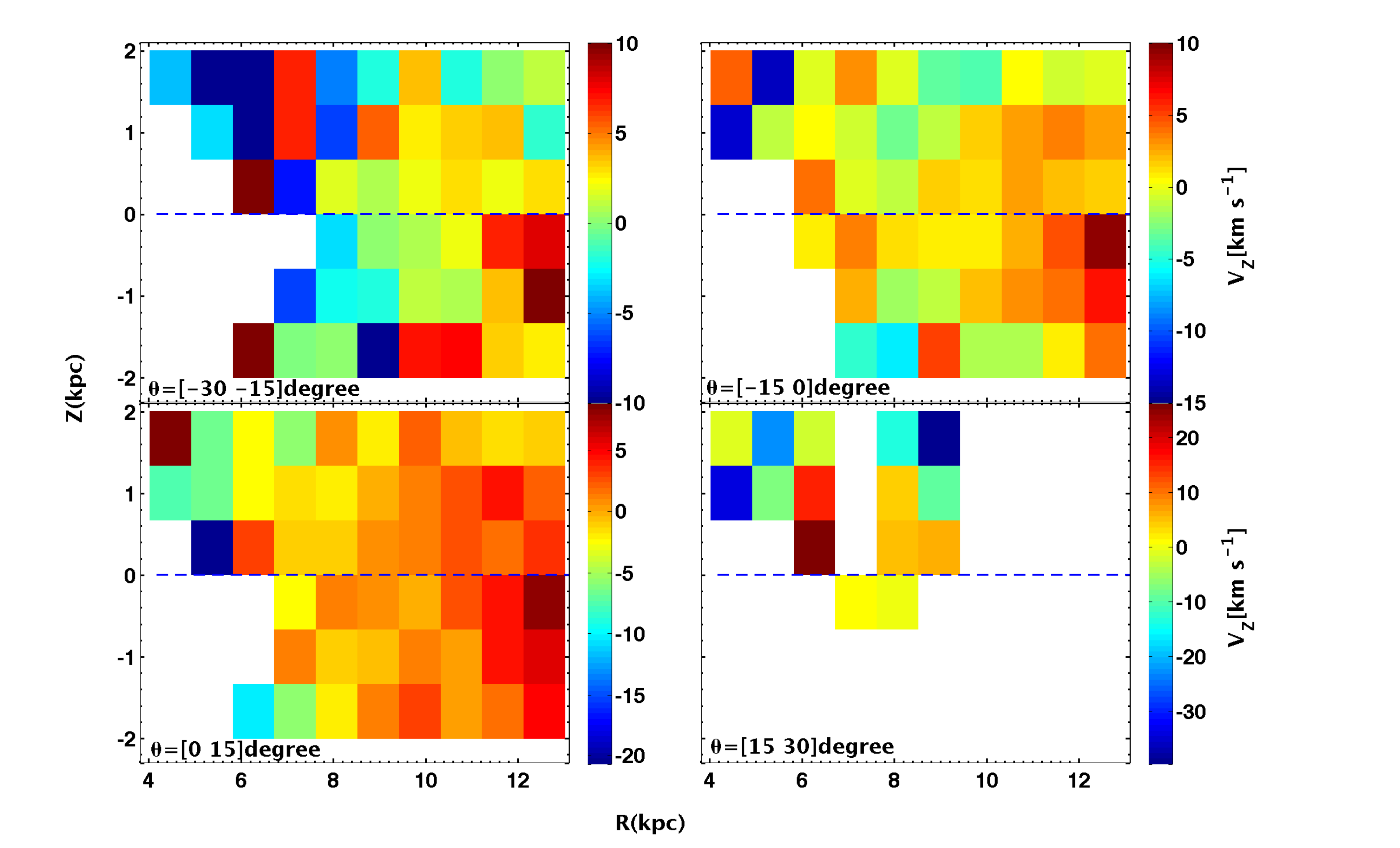}
  \caption{Edge-on maps of the median vertical velocity $V_Z$ of the red clump sample. Each map corresponds to a slice of 15 degrees in azimuth: [-30, -15] (top left), [-15, 0] (top right), [0, +15] (bottom left), and [+15, +30] degrees (bottom right). The Sun is at X = 8.34 \,kpc and Y = 0\,kpc.}
  \label{vz_projections}
\end{figure*}

Similarly, we also consider the projections of vertical velocity. In Figure~\ref{vz_projections}, we can see very clear features of the bending mode or vertical bulk upward motions in the range [-15 15] degrees, this range has high sampling in the LAMOST survey. The stars show significant upward motions from $V_Z \sim $ 1 km s$^{-1}$ to 15 km s$^{-1}$, especially between $-1 < Z < 1$ \,kpc.  Some possible mechanisms to excite these motions are discussed in the next section. 

\begin{figure*}
  \centering
  \includegraphics[width=0.98\textwidth]{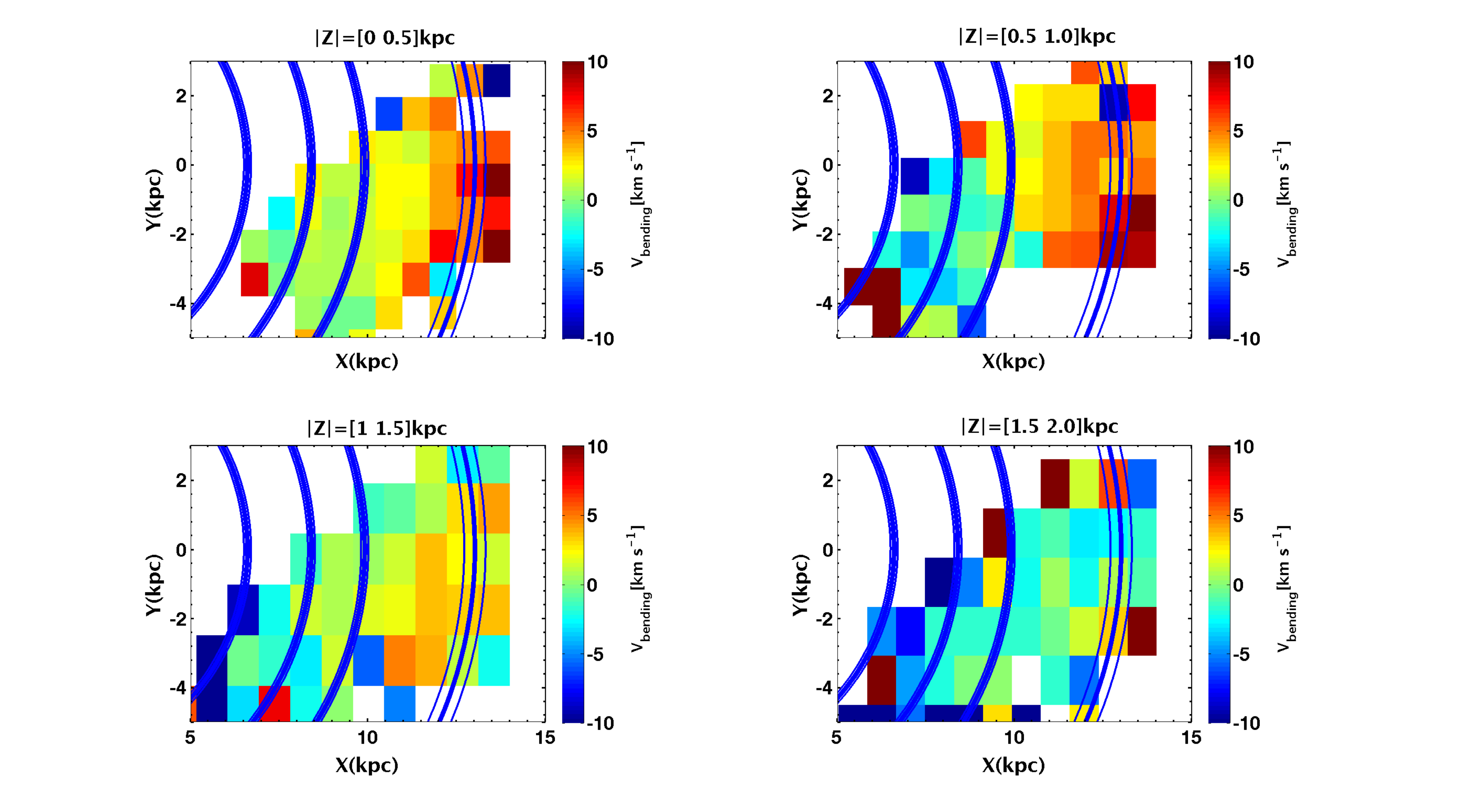}
  \caption{The figure shows the bending mode's contribution to bulk velocities. Here the disk has been divided into four groups of symmetric layers according to the \citet{Katz2018} formula. The distance to the Galactic mid-plane increases from top to bottom: [-0.5, 0] and [0, 0.5] \.kpc (top left), [0.5, 1.0] and [-1, -0.5] \,kpc (top right), [-1.5, -1.0] and [1, 1.5] \,kpc (bottom left) and [1.5, 2.0] and [-2.0, -1.5] \,kpc (bottom right). The general trend of the bending mode is increasing with radius. The blue bold solid lines and slim lines are spiral arms from \citet{Reid14} }.
  \label{quantityben}
\end{figure*}

\begin{figure*}
  \centering
  \includegraphics[width=0.98\textwidth]{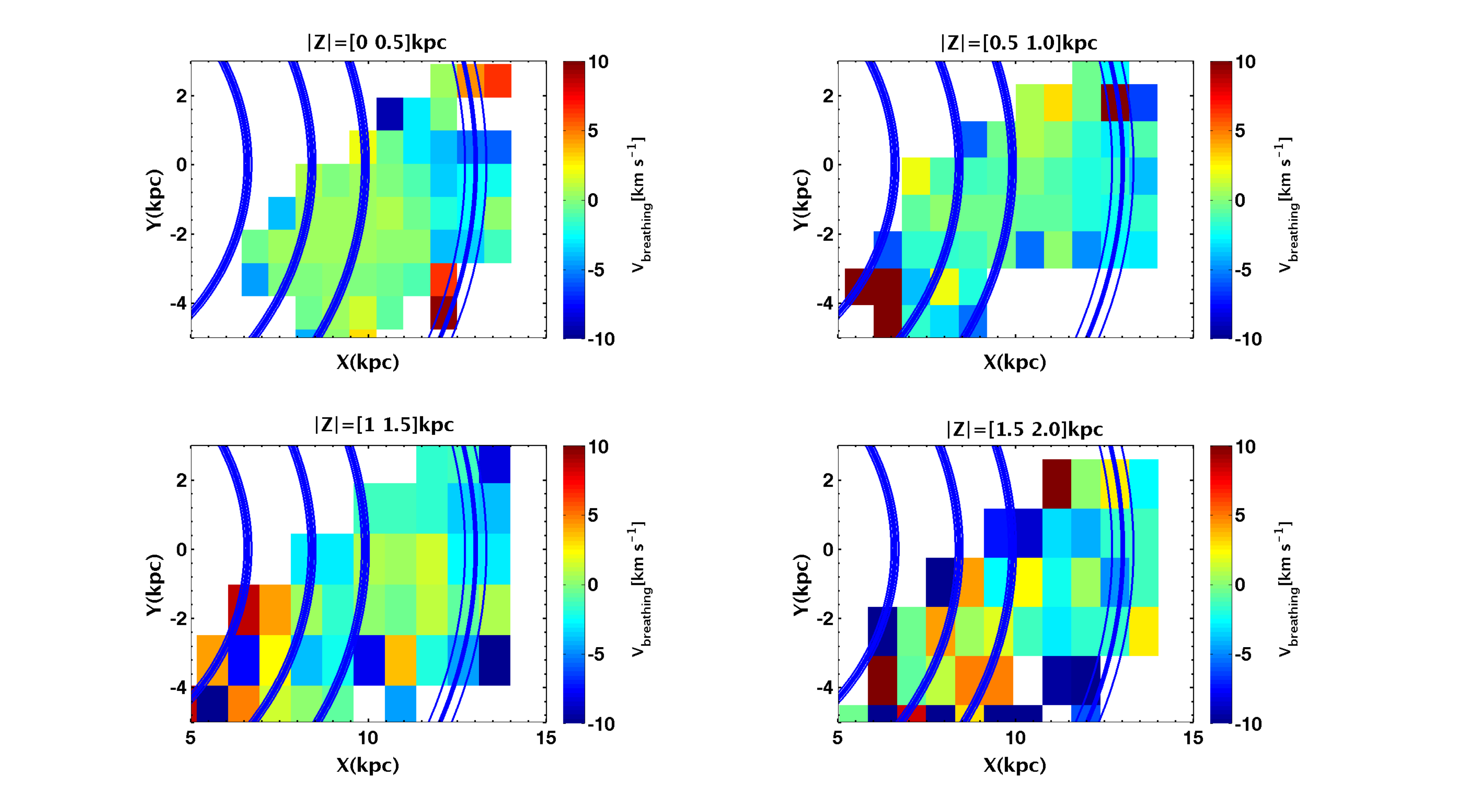}
  \caption{The figure shows the breathing mode's contribution to bulk velocities. Here the disk has been divided into four groups of symmetric layers according to the \citet{Katz2018} formula. The distance to the Galactic mid-plane increases from top to bottom: [-0.5, 0] and [0, 0.5] \.kpc (top left), [0.5, 1.0] and [-1, -0.5] \,kpc (top right), [-1.5, -1.0] and [1, 1.5] \,kpc (bottom left) and [1.5, 2.0] and [-2.0, -1.5] \,kpc (bottom right). The general trend of the breathing mode is decreasing with radius.}
  \label{quantitybrea}
\end{figure*}

In order to quantify the bending and breathing mode contributions, we also map the bending and breathing velocities according to \citet{Katz2018} formula:
\begin{equation}
V_{bending}(X,Y) = 0.5\ [\tilde{V}_Z((X,Y), L) + \tilde{V}_Z((X,Y), -L)]
\label{eq:bend}
\end{equation}
and
\begin{equation}
V_{breathing}(X,Y) = 0.5\ [\tilde{V}_Z((X,Y), L) - \tilde{V}_Z((X,Y), -L)]
\label{eq:breath}
,\end{equation}
where $L\equiv Z$. We choose the symmetric layer $L$ in the north Galactic hemisphere and in the south Galactic hemisphere, then calculate the median velocity in the cell $(X,Y)$ and in the horizontal layer $L$: $\tilde{V}_Z((X,Y), L)$, and use these to get the relative contributions of both modes. Figure~\ref{quantityben} and Figure~\ref{quantitybrea} show the bending mode and breathing mode contributions on the X-Y plane at different symmetric heights, with color representing the velocity value. In this work, the disk has been divided into four groups of symmetric layers according to the \citet{Katz2018} formula. The distance to the Galactic mid-plane increases from top to bottom: [-0.5, 0] and [0, 0.5] \,kpc (top left), [0.5, 1.0] and [-1, -0.5] \,kpc (top right), [-1.5, -1.0] and [1, 1.5] \,kpc (bottom left) and [1.5, 2.0] and [-2.0, -1.5] \,kpc (bottom right). It is clear that the general trend of breathing is decreasing with radius outwards in the X,Y plane while bending is increasing in the same direction. The blue lines show the four spiral arms for reference \citep{Reid14}.

\subsection{1$-$D velocity asymmetric structure along the radial direction}

\begin{figure*}
  \centering
  \includegraphics[width=0.98\textwidth]{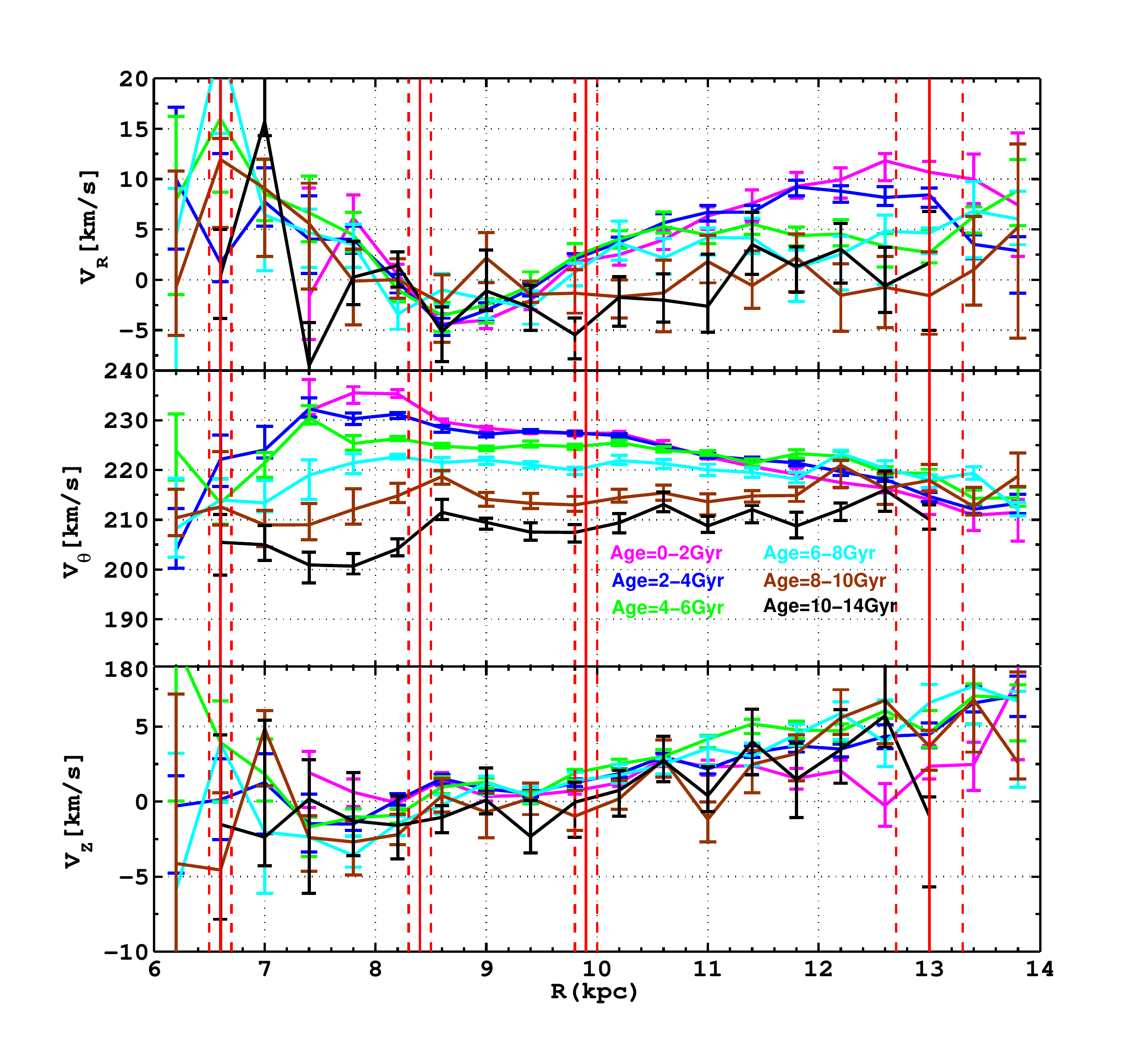}
  \caption{Median velocity of stars between Z=[-1, 1] \,kpc, toward the Galactic anticenter. The top panel shows the median Galactocentric radial velocities, $V_{R}$, of the red clump giant sample as a function of Galactic radius in different ages: [0,2] \,Gyr (cyan) to [10-14] \,Gyr (black). The error bars or uncertainties on the median radial velocities are from a bootstrap method. The middle panel shows the variation of median $V_{\theta}$, which ranges from 200 km~s$^{-1}$ increasing to 235 km~s$^{-1}$ . The younger stars exhibit larger median $V_{\theta}$ than older ones with radius at least less than 12 \,kpc. The bottom panel shows the vertical velocity $V_{Z}$, along R. There is a trend from negative to positive with upward bulk motions at all radii R$>$9 kpc for all populations, with no prominent difference for each populations in the general trends. The red vertical solid and dashed lines in all panels mark the locations of spiral arms \citep{Reid14}.}
  \label{vrvzvphi_R}
\end{figure*}

The top panel of Figure~\ref{vrvzvphi_R} shows the median radial velocity, $V_{R}$, as a function of Galactocentric radius $R$ in different age populations; different colors represent different ages from [0-2], [2-4], [4-6], [8-10], [10-14] \,Gyr. In this figure, the median radial velocity has a negative slope in the inner regions, and a positive trend in outer parts, with the turning point around $R = 8.5$\,kpc. This is similar to the U-shape found by \citet{Katz2018}, with the negative trend meaning that more stars move inwards than outwards. Beyond 9 \,kpc, the median radial velocity becomes positive, meaning that more stars move outwards than inwards, in agreement with \citet{Lopez2018} and \citet{Lopez2019}. There is a dip around the Sun's location, which is similar to the results by \citet{Wang2018a}. It is not very likely to be a artificial feature due to the systematic errors for the precise red clump star distances and proper motions. Solid lines with error bars represent the median radial velocity of stars between Z=[-1, 1] \,kpc, because we do not want to consider more thick disk population and most of our red clump stars are in this range. Apart from this, we also want to avoid the influence of moving groups or streams on our discussion.  The U-shape feature has a different pattern in different age bins. We can see the velocity of younger populations (0-4\,Gyr) is larger than that for older populations. The radial velocity values are from around $-$5 km~s$^{-1}$ to 10 km~s$^{-1}$ for this panel. For reference, the red vertical solid and dashed lines in all panels mark the locations of spiral arms \citep{Reid14}.

The middle panel displays the variation of median $V_{\theta}$ with $R$. The value ranges are  from $\sim$200 km s$^{-1} $ to 230 km s$ ^{-1}$ for stars of different ages. There are clear differences between the younger ones and older ones, showing that the younger stars have higher velocities than old stars, some of which roughly corresponds to the results in \citet{Wang2018a}. This reflects the rotational velocity gradient with height from the midplane; older stars have larger velocity dispersion and larger asymmetric drift, so the younger stars show a larger mean rotational velocity than older ones. At distances larger than 12\,kpc, these trends mix together because the region is far away from the bar and spiral arms, and stars of the outer disk are generally relatively younger compared with inner disk so that the velocity difference is not so large. 

The bottom panel shows the vertical velocity $V_{Z}$ along $R$. There is a clear trend with radius from negative (downward) to positive (upward) bulk motions. Furthermore, the vertical velocities of the younger stars do not show much difference with older populations due to the possible age accuracy, except that the youngest population (0-2)\,Gyr has a decreasing trend around 13\,kpc. Some possible scenarios to explain the trends noted here are discussed in the next section. Please notice that we only concentrate on stars less than around 13-14\,kpc.

\section{Discussion: Possible implications and comparisons}

\subsection{Discussion of radial motions}

\subsubsection{Northern structure in radial motions}
\citet{Xu15} revealed there is a northern overdensity around 2-4 \,kpc outside of the Sun. \citet{Wang2018b, Wang2018c} confirmed this substructure by using LAMOST red giant branch stars. Results found by \citet{Tian172} showed that there are positive radial motions beyond 9\,kpc. This gradient might be caused by the northern substructure, but \citet{Tian172} detected the 1D asymmetrical radial motions in roughly two age populations by integrating stars on both sides. In \citet{Wang2018a} we also see 2D asymmetrical motions including the north near velocity substructure and 1D radial motions with a positive gradient similar to \citet{Tian172}. Moreover, in \citet{Wang2018a} we can also see clearly the northern velocity contributing to these positive increases. However, in this work, we present a full map and give time stamps on this structure for the first time.

\begin{figure*}
  \centering
  \includegraphics[width=1.1\textwidth]{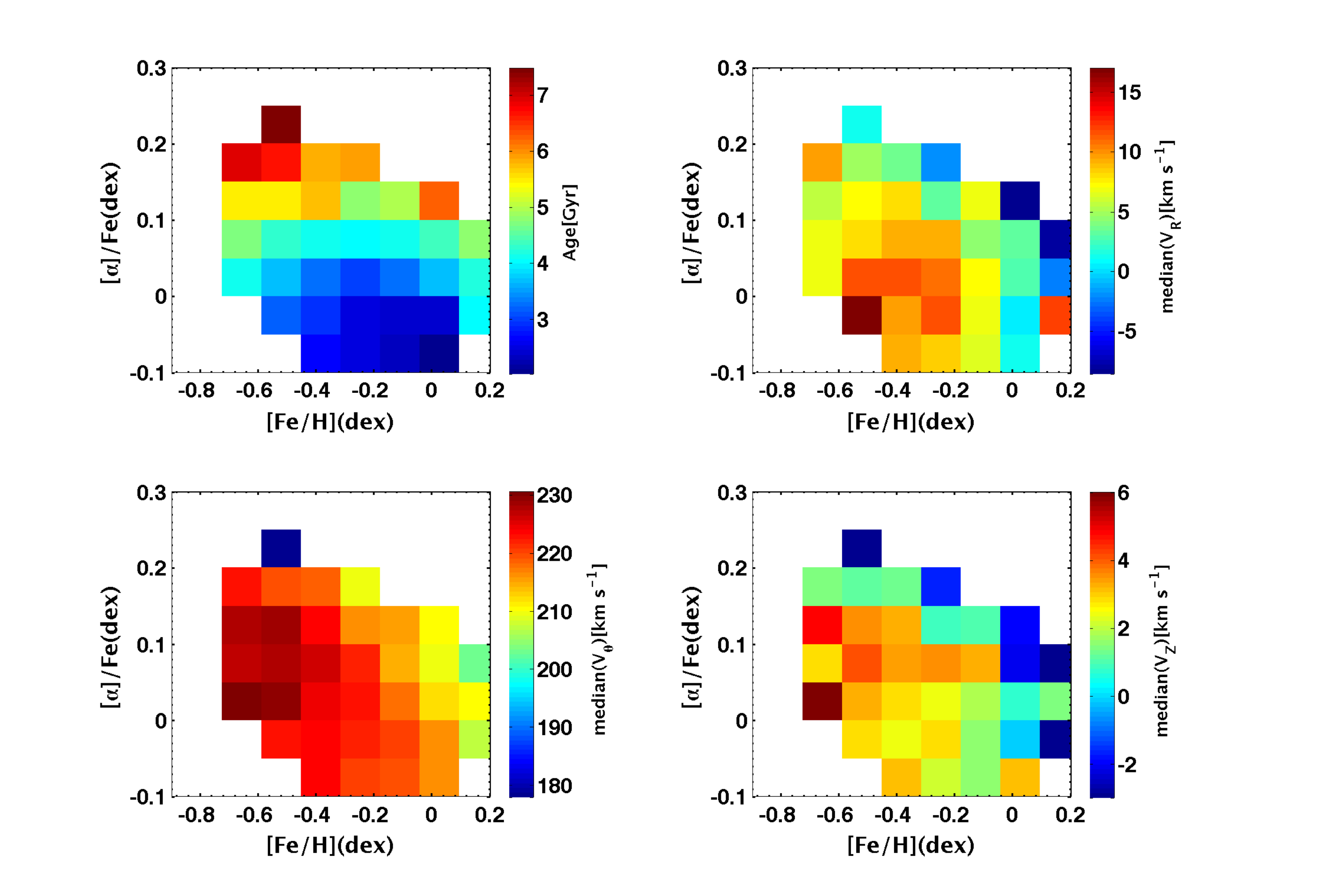}
  \caption{ The top left panel shows the age distribution for red clump giant stars on the [Fe/H] and [$\alpha$/Fe] plane. The color-coded map shows the age distribution for the sample less than 15 \,Gyr, Z=[0.2 0.8] \,kpc, R=[10-13] \,kpc. The top right panel is the radial velocity distribution for red clump giant stars on the [Fe/H] and [$\alpha$/Fe] plane. Similarly, the bottom left one is rotational velocity, and the bottom right one is the vertical velocity distribution for the rough north near region in this work.}
  \label{rcfehafevrvzvphi}
\end{figure*}

In order to see more chemo-kinematical details of this northern structure, we can see in Fig.~\ref{rcfehafevrvzvphi} (top-left panel) the age distribution for red clump giant stars on the [Fe/H] and [$\alpha$/Fe] plane. The color-coded map shows the age distribution for a sample with age less than 15 \,Gyr, Z=[0.2 0.8] \,kpc, R=[10-13] \,kpc. The top right panel is the radial velocity distribution for red clump giant stars on the [Fe/H] and [$\alpha$/Fe] plane. Similarly, the bottom two panels correspond to rotational velocity and vertical velocity distributions for the rough north near region in this work separately. As we can see, the overall trend for this structure's population is present in stars younger than 6 \,Gyr, more metal rich than 0.7 \,dex, and with [$\alpha$/Fe] lower than 0.2 \,dex (top left). This population of stars moves outward on average with radial velocity larger than zero (Fig.~\ref{rcfehafevrvzvphi}, top-right panel) and has large rotational velocity. It is worth mentioning that Fig.~\ref{rcfehafevrvzvphi} (lower left panel) shows the more metal-poor stars of the thin disk have higher angular momentum. This has been attributed to these more metal-poor stars migrating inwards from the more metal-poor outer disk where the angular momentum is higher; this effect was also found by \citet{Lee2011}. It is interesting to see that the effect is still present at R = 10-13 \,kpc.

\subsubsection{Some possible scenarios for radial motions}
For in-plane asymmetries, large-scale non-circular streaming motions would clearly be due to a non-axisymmetric component of the Galactic potential, which could be caused by the Galactic bar, spiral arms, inclined orbits due to a warp or its motion, a triaxial dark matter halo \citep{Siebert11, Siebert12}, or intrinsically elliptical or a net secular expansion of the disk \citep{Lop16}. It is likely contributed by a combination of some or all of these components \citep{Wang2018a}. From the top panel in Fig.~\ref{vrvzvphi_R}, we can also see that these possible effects have larger dynamical effects on the younger populations than older ones. We can also see the effects might strongly affect the populations with ages between 0-6 \,Gyr.

\subsubsection{Spiral arms}


\citet{Kawata2018} found many diagonal ridge features and proposed some of these are likely related to the perturbations from the bar's outer Lindblad resonance and spiral arm dynamics. We can see there is a dip in  Fig.~\ref{vrvzvphi_R} (top panel), around 8.5\,kpc: it corresponds to the local spiral arm location. We speculate that in-plane asymmetries might also be contributed by spiral arms in the solar neighborhood. For the outer disk, radial velocity asymmetries might be produced by the spiral structures that are always spatially correlated with spiral arms \citep{Siebert12, Faure14, Grand15}. The analysis of the 3D response of the disk distribution function to a spiral arm perturbation can also be found in \citet{Monari16}.

If we investigate in more detail, we can see the radial velocity is from around  -5 km s$^{-1}$ to 10 km s$^{-1}$ for the populations younger than 4-6 \,Gyr, with a gradient of about 3 km s$^{-1}$ kpc $^{-1}$. For the populations older than 4 \,Gyr, the value is generally from around  -5 km s$^{-1}$ to 5 km s$^{-1}$ corresponding to a gradient of around 2 km s$^{-1}$ kpc $^{-1}$. Recently, \citet{Lopez2019} have analyzed the radial velocity from 8$-$28 \,kpc with APOGEE \citep{Majewski2017} data. We find our results are very similar to their Figure 2 results. \citet{Lopez2019} think factors such as the effect of the Galactic bar, streams, or mergers do not seem appropriate to explain their observations. A possible explanation proposed by them might be the gravitational attraction of overdensities in a spiral arm, with compression where the stars enter the spiral arm and expansion where they exit, corresponding to negative and positive values. We find for our results, if we choose the mean value of different populations of radial velocity, it is about 6 km s$^{-1}$. According to \citet{Lopez2019}, the mass in spiral arms necessary to produce these velocities would be around 3\% of the mass of the disk, consistent with current knowledge of the spiral arms. Therefore, we think that the spiral arms may explain
our kinematical data.

 \subsubsection{A simple class of out-of-equilibrium systems }
Another possible scenario that \cite{Lopez2019} explore is a simple class of systems in which radial motions are generally created by the monolithic collapse of isolated self-gravitating overdensities. In this simple out-of-equilibrium system, a quasi-planar spiral structure surrounding a virialized core will be produced by rotating and asymmetrical mass distributions that evolve under their own gravity.
This model predicts non-circular orbits in the outer disk. Mean orbits in the very outer disk are out of equilibrium, so they have not reached circularity yet. A galaxy in equilibrium does not need orbits to be circular and indeed orbits of individual stars in galaxies are generally not circular in equilibrium axisymmetric configurations; but it is the ``mean orbit'' that is circular in the axisymmetric potential. However, in a non-equilibrium regime there is a net average non-zero radial velocity. Under some initial conditions, they are able to reproduce the observed features. See more details in \citet{Benhaiem2017, Lopez2019, Lopez2019b}. We also think the out of equilibrium model might be a possible scenario to explain our data.
 

\subsubsection{Minor mergers}
Minor mergers mainly contribute to vertical waves. These perturbations were thought not to affect the in-plane radial velocity \citep{Gomez13,Tian172}. However, \citet{Carrillo2019} have more recently shown that a major perturbation, such as the impact of Sagittarius, could reproduce a radial asymmetrical velocity field in their observations too. In fact, from the most recent simulations, the bending mode induced has a strong radial velocity counterpart, as can be seen for example in Figure 4 of  \citet{Laporte2019}. Therefore, a minor merger might also be contributing to these asymmetries according to the newest simulations and data.

 \subsubsection{Bars, intrinsically elliptical outer disk, secular expansion of the disk, streams}
 
The outer disk asymmetric radial motions are likely mainly contributed by the bar dynamics with a given pattern speed \citep{Grand15, Monari15, Tian172, Liu173}. \citet{Tian172} find that the mean radial velocity is negative within R $\sim$ 9 \,kpc and positive beyond, which might imply a perturbation induced by the rotating bar with pattern speed of 45 km s$^{-1} $.  A simulation by \citet{Liu173}  showed that the bar pattern speed of 60 km s$^{-1}$ can be matched with their data,  although other  recent works 
give lower values around 40 km s$^{-1}$ \citep{Portail2017, Bovy2019, Sanders2019}. With the help  of Schwarzschild's orbit-superposition technique, \citet{Wangyg2012} constructed self-consistent models of the Galactic bar and found that the best-fitting Galactic bar model has a pattern speed $\omega_{p}$ = 60 km s$^{-1}$, which also implies that the bar might be in the plane and have angle on the X-Y plane. However, the asymmetrical variation with age on the R-Z plane seen in Fig.~\ref{vR_age_benbrea} mainly happens on the northern side, and we do not understand how a bar might produce this asymmetry. Moreover, \citet{Monari14} showed that no gradient in the radial velocity is expected from a bar effect for the observed distances. For the current understanding, we suppose that the bar might not be the main contributor.

\citet{Lop16} and \citet{Lopez2019} propose that in plane asymmetries might be explained by the mean orbits of disk stars being intrinsically elliptical, or a perturbation due to net secular expansion of the disk. In the case of elliptical orbits, radial motions would indicate an eccentricity, $e$, different from zero, but similar distributions in their works' calculations find small $e$, not significantly different from zero. If we were in the case of secular expansion, the lifetime of the Galaxy would be much longer than their calculation and the size of the Galaxy cannot change so fast. We think these two scenarios are not  likely.

A local stream cannot be the explanation for radial velocities along a wide range of a few \,kpc. Moreover, a large-scale stream associated with the Galaxy in the Sun-Galactic center line should have some additional clear evidence of it embedded in our Galaxy. We don't detect any signatures, so we think this is not very likely.  According to \citet{Lopez2019}, who used a similar analysis but divided their sample into the north and south side, they found that there is a relatively small asymmetry between the northern and southern Galactic hemispheres. For the same region, we calculate the difference for both sides, and find that the difference is around 0.5$-$5 km s$^{-1}$. We suggest that it is unlikely that the same stream contributed by an infalling galaxy would be so symmetric with respect to the mid-plane.

In summary, we think the in-plane asymmetries are not likely mainly contributed by the bar, an intrinsically elliptical outer disk, secular expansion of the disk, or streams. We propose that spiral arms dynamics, out-of-equilibrium models, minor mergers are the important contributors. Dark matter sub-halos, warp dynamics or other mechanisms might also be possible which is needed to be investigated in the future.

\subsection{Discussion for vertical motions}
The overall trends for vertical motions in different populations of this work are similar to recent work by \citet{Lopez2018}.  We see that different populations have similar patterns with radius in Fig.~\ref{vrvzvphi_R}, and we can see the bending mode exists from 0$-$8\,Gyr in Fig.~\ref{vz_age_benbrea}. The scenarios to excite vertical bulk motions must be more complex, including scenarios such as perturbations due to the passing of the Sagittarius dwarf galaxy proposed by \citet{Gomez13} or the LMC \citep{Laporte18}, or the disk response to bombardment by merging lower-mass satellites \citep{Donghia16}. The effects of even lower-mass dark matter subhalos have also been invoked as  a possible explanation \citep{Widrow14}.  

The stellar disk has a clear warp \citep{Lop02}. According to simple analyses of vertical velocities \citep{roskar10,Lop141}, the Galactic warp's line-of-nodes is located close to the line Galactic anticenter direction \citep{Chen2019}. By assuming that this vertical motion is the result of the warp modeled as a set of circular rings that are rotated and whose orbit is in a plane with angle with respect to the Galactic plane, we calculate its contribution to the vertical velocity. A simplified vertical motion model \citep{Lop141}
includes two terms for the vertical velocity, a first one for the inclination of the orbits and the second from the temporal 
variation of the amplitude of the warp:

\begin{equation}
V_{Z}(R>R_\odot)\approx \frac{(R-R_\odot )^\alpha}{R}[\gamma \omega_{LSR}\cos (\phi -\phi _w)
\end{equation}\[
\hspace{3cm} +\dot{\gamma }R\sin (\phi -\phi _w)].\]
where $\phi _w$ is the azimuth of the line of nodes, $\gamma $  is the amplitude of the warp and $\dot{\gamma }$ describes the warp amplitude evolution. By assuming  $\alpha =1$ \citep{Reyle2009}, $\phi _w=5$ deg (in the literature the values are between -5 and +15 deg; \citealt{Momany2006, Reyle2009}), we just use the data with $R\ge 8.34$ kpc to get the best fitting value based on Markov chain Monte Carlo (MCMC) simulation provided by EMCEE \citep{Foreman$-$Mackey2013}.

The values of $\gamma $ and $\dot{\gamma }/
\gamma $ that fit our data are
\begin{equation}
\label{gamma}
\gamma =-0.05\pm 0.01
,\end{equation}
\begin{equation}
\frac{\dot{\gamma }}{\gamma }=11.25\pm 1.27\,{\rm Gyr}^{-1}
\label{gammadot}
.\end{equation}

\begin{figure}
  \centering
  \includegraphics[width=0.48\textwidth]{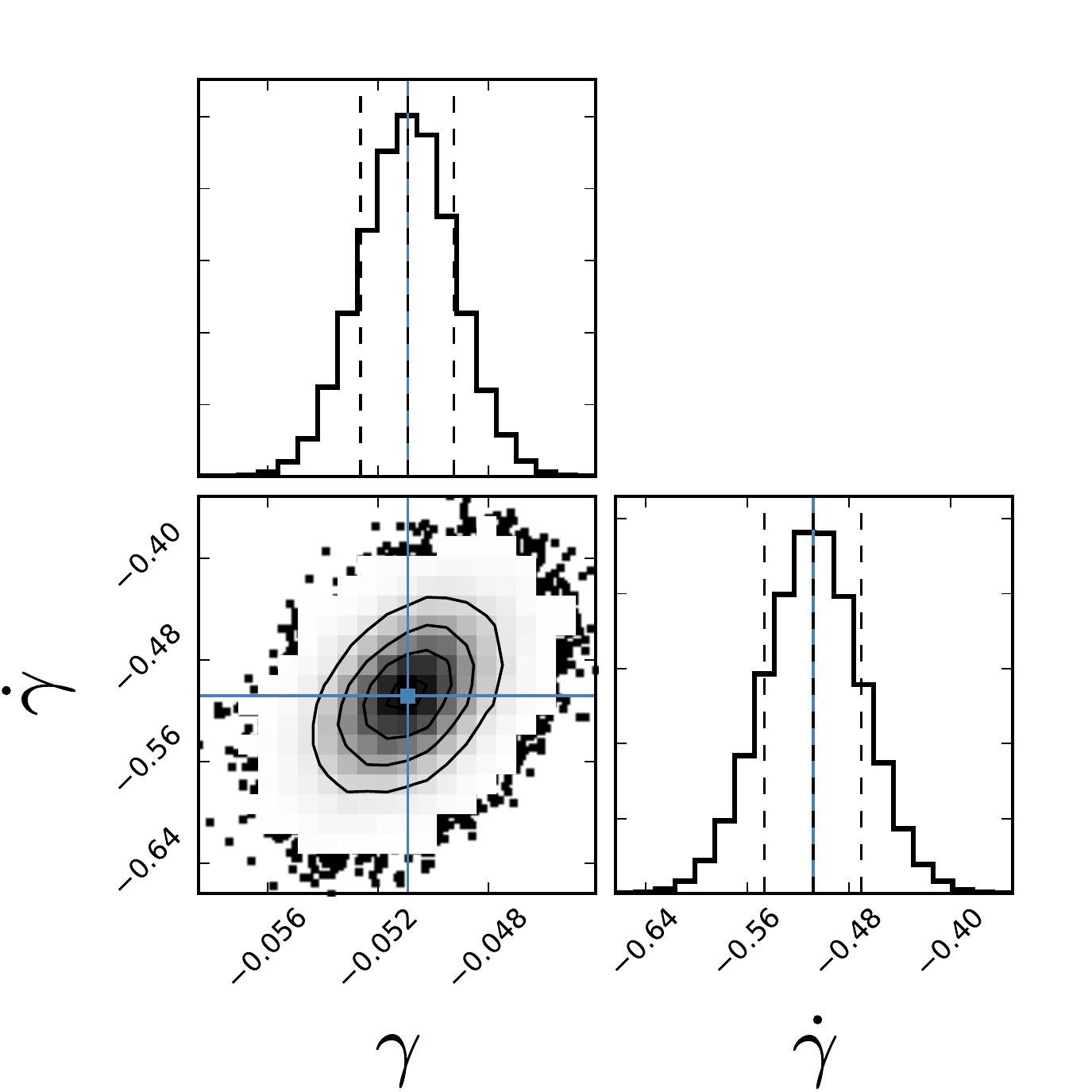}
  \caption{ The likelihood distribution of the parameters ($\gamma $  and $\dot{\gamma }$ ) drawn from the MCMC simulation, The solid lines in the histogram panels indicate the maximum likelihood values of the parameters. The dashed lines indicate the 1$-\sigma$ regions defined by the covariance matrix.}
  \label{warpfitting}
\end{figure}

Eq. (\ref{gammadot}) indicates that our warp might not be stationary ($\dot{\gamma }=0$) if we assume the model is right. Our results are different from the 0.23$\pm$ 0.16 of $\gamma $ and $\dot{\gamma }/\gamma =-34\pm 17\,{\rm Gyr}^{-1}$ in \citet{Lop141}; the reason might be that the contribution to $\dot{\gamma }$ comes from the southern warp and is negligible in the north in their work, but for LAMOST, we can only be sensitive to the northern warp due to low sampling and coverage in the southern sky. Moreover, \citet{Lop141} used data from PPMXL \citep{Roeser10} that had much larger error bars in proper motions than Gaia-DR2, so that their fit was of poorer quality. The fitting result is shown in Figure~\ref{warpfitting}.

\begin{figure}
  \centering
  \includegraphics[width=0.5\textwidth]{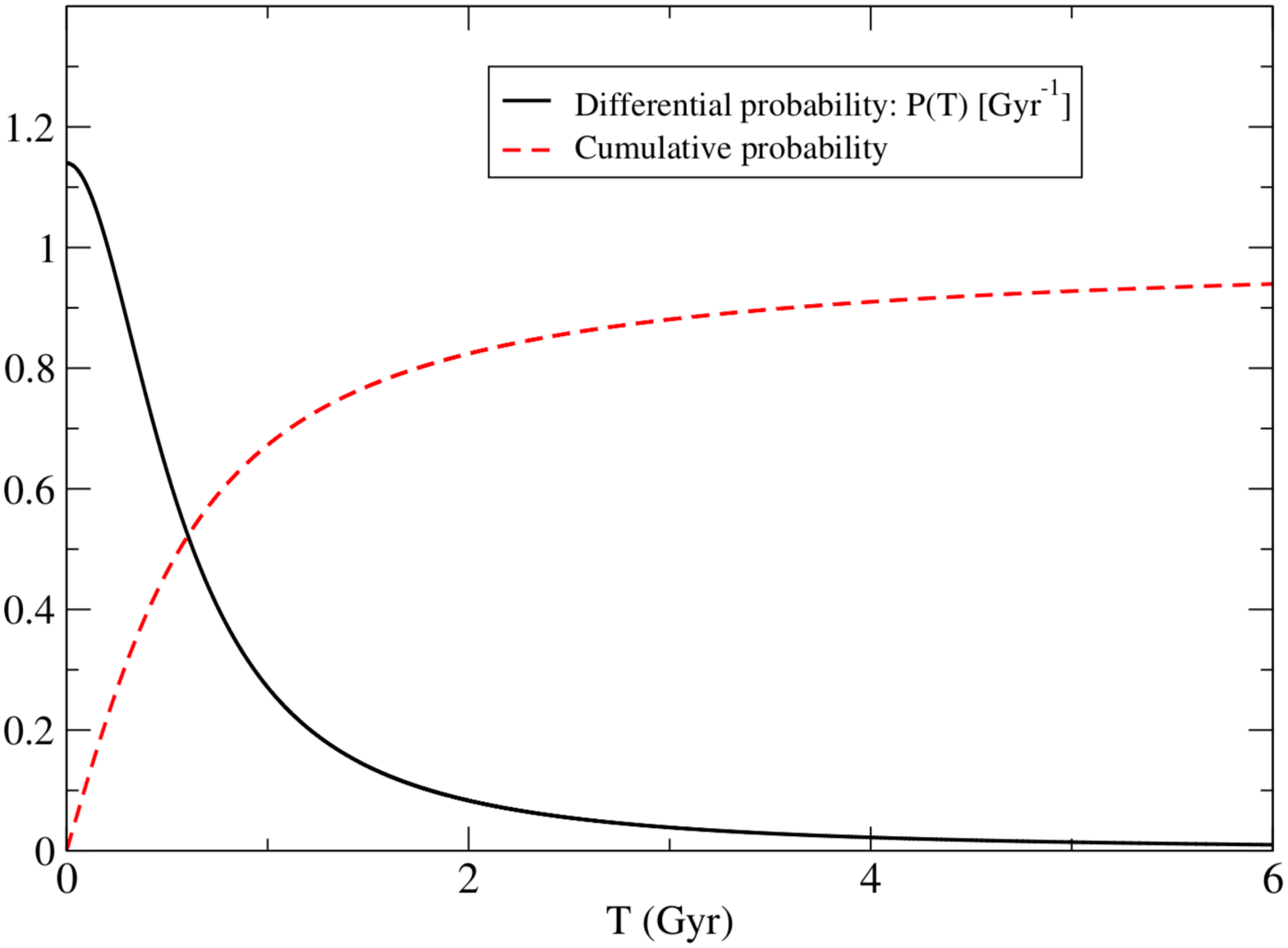}
  \caption{Distribution of probability of the period for the motion of $\gamma (t)=\gamma _{\rm max}\sin (\omega t)$ given by Eq. (\ref{gammadot}).}
  \label{warpperiod}
\end{figure}

According to current results, we can also attempt to  derive some information of the warp kinematics. Eq. (\ref{gammadot}) indicates that our warp is not stationary ($\dot{\gamma }=0$). If we assume a sinusoidal oscillation, $\gamma (t)=\gamma _{\rm max}\sin (\omega t)$, we have a period
\begin{equation}
T=\frac{2\pi }{\omega }=2\pi \left(\frac{\dot{\gamma }}{\gamma }\right)^{-1}\cot (\omega t)
\label{t}
,\end{equation}
and the probability of having a period $T$ is the convolution of two probability distributions \citep{Lop141} [equation 19]:
\begin{equation}
P(T)dT=\frac{dT}{2^{1/2}\pi^{5/2}\sigma _x}\int _{-\infty}^{+\infty} dx \frac{|x|}{1+\left(\frac{Tx}{2\pi }
\right)^2}e^{-\frac{(x-x_0)^2}{2\sigma _x^2}}
,\end{equation}
where $x_0\equiv \frac{\dot {\gamma}}{\gamma}$ and $\sigma _x$ is its r.m.s.  Figure~\ref{warpperiod} shows this probability distribution. From this distribution, 
the cumulative probabilities of 0.159, 0.500, and 0.841 are given for $T=0.143$, 0.563, and 2.225 \,Gyr, respectively, so we can say that $T=0.56^{1.66}_{-0.42}$ Gyr (68.3\% C.L.),
or $T=0.56^{+11.27}_{-0.54}$ Gyr (95.4\% C.L.). Alternatively, we can say that $T<1.02$ Gyr (68.3\% C.L.), $T<6.72$ \,Gyr (95.4\% C.L.).

Our results are equivalent to a rotation of the rings around the line of nodes with an angular velocity of $-$0.56$\left(1-\frac{R_\odot }{R}\right)$ km/s/kpc, which is less than  0.2 km/s/kpc in the range of $R$ between 8 and 13 \,kpc.

As we can see in Figure~\ref{quantityben} and Figure~\ref{quantitybrea}, our general trends are consistent with the simple ``bending'' and ``breathing'' model, so we speculate that the vertical bulk motions are dominated by these modes.  The Galactic bar and spiral arms can induce breathing modes, and external perturbations can induce bending modes  \citep{Faure14, Monari15,  Monari16}. \citet{Chequers18} found that bending waves could also arise without excitation by a satellite/merging event. \citet{Widrow14} used a toy-model simulation of disc-satellite interactions to show that a passing satellite galaxy could produce bending or breathing modes depending on the vertical velocity of the satellite as it passes through the Galactic disc. The picture can actually be rather complex;  \citet{Carrillo17} mentioned that the external perturbations also generate internal spiral perturbations which can in turn excite breathing modes. There is a possibility that the bars, spiral arms, minor mergers, warps, bending and breathing modes are coupled together \citep{Widrow2019}.

Minor merger simulations carried out by \citet{Gomez13} implied that  variations of vertical motions less than $\sim10$ km s$ ^{-1} $ and radial and azimuthal variations of the mean vertical velocity will correlate with the spatial structure if vertical motions are caused by spiral arms, as is shown in the bottom panel of Fig.~\ref{vrvzvphi_R}. There is no clear spatial correlation, so the vertical motions in the outer disk are not likely mainly contributed by spiral arms. The Galactic bar is unlikely to induce mean vertical motions greater than $\sim$ 0.5 km s$^{-1}$ in the outer disk \citep{Monari15}. Our measured bulk motions are larger than 0.5 km s$^{-1} $, so we do not favor a bar/spiral arm as main contributors for the vertical asymmetric motions, although they might be  possible factors.

In summary, according to current analysis, the complex bending and breathing modes produced by the internal perturbation or minor merger or other external perturbation is the most possible scenario. Warp contributions or other mechanisms might also be a reasonable possible explanation. Possibly they are coupled together to cause the vertical asymmetry with bending and breathing modes accompanied with mean non zero radial motions for the regions studied in this work.

\section{Conclusions}

In this work, we investigate the kinematics of mono-age populations of red clump giant stars in the Galactic disk between $R$ = 6 to 14 \,kpc, $Z$ = -2 to 2 \,kpc with the recent LAMOST-DR4 and Gaia proper motion catalogue. From time stamps on the asymmetrical variations of the velocities, especially for Galactocentric radial velocity and vertical velocity, not only do we confirm Gaia and LAMOST results, but we also find that:

The median radial velocity substructure located at $Z\sim$ 0.5 \,kpc and $R\sim$ 9$-$12 \,kpc corresponds to the northern overdensity (north near structure), which is sensitive to the perturbations in the early 6 \,Gyr. The radial velocity for younger populations is larger than older ones. 

The stars beyond $R \gtrsim 9$~kpc are moving upward on average in different age populations. We investigate the temporal evolution of this structure.  It appears that the bending mode vertical motions sensitive duration is around 8 \,Gyr.  We also simply show that if the vertical motions are contributed by the warp, then it implies the warp is a not stationary structure.
  
With the help of previous works, we discuss that in-plane asymmetries are  not likely mainly contributed by the bar, intrinsically elliptical outer disk, secular expansion of the disk, or streams. The gravitational attraction of overdensities in a spiral arm or arm dynamics, the simple class of out-of-equilibrium systems, or minor mergers might be important factors, although we cannot exclude other scenarios such as dark matter sub-halos, warps, etc. For the vertical asymmetries, we can rule out that they might be mainly caused by spiral arms or simply by the bar. Other mechanisms such as minor mergers, warp line of nodes, dark matter sub-halos, or satellite accretion might also contribute, leaving an imprint in the bending and breathing modes that we observe. The most possible scenario is that all these mechanisms are coupled together to cause in-plane and vertical asymmetric features simultaneously.

\section*{Acknowledgements}
We would like to thank the anonymous referee for his/her helpful comments. We also thank Lawrence M. Widrow, Heidi J. Newberg, Ivan Minchev and Wang Y.G. for helpful discussions and comments. HFW is supported by the LAMOST Fellow project, National Key Basic Research Program of China via SQ2019YFA040021 and funded by China Postdoctoral Science Foundation via grant 2019M653504 and Yunnan province postdoctoral Directed culture fundation. MLC was supported by grant PGC- 2018-102249-B-100 of the Spanish Ministry of Economy and Competitiveness (MINECO). JLC acknowledges support from the U.S. National Science Foundation via grant AST-1816196. Y.H. acknowledges the National Natural Science Foundation of China U1531244,11833006, 11811530289, U1731108, 11803029, and 11903027 and the Yunnan University grant No.C176220100006 and C176220100007. Guoshoujing Telescope (the Large Sky Area Multi-Object Fiber Spectroscopic Telescope, LAMOST) is a National Major Scientific Project built by the Chinese Academy of Sciences. Funding for the project has been provided by the National Development and Reform Commission. LAMOST is operated and managed by the National Astronomical Observatories, Chinese Academy of Sciences. The LAMOST FELLOWSHIP is supported by Special Funding for Advanced Users, budgeted and administrated by Center for Astronomical Mega-Science, Chinese Academy of Sciences (CAMS). This work has also made use of data from the European Space Agency (ESA) mission{\it Gaia} (\url{https://www.cosmos.esa.int/gaia}), processed by the {\it Gaia} Data Processing and Analysis Consortium (DPAC, \url{https://www.cosmos.esa.int/web/gaia/dpac/consortium}). Funding for the DPAC has been provided by national institutions, in particular the institutions
participating in the {\it Gaia} Multilateral Agreement.

\bsp	
\label{lastpage}
\end{document}